\documentclass[reprint,aps,showpacs,showkeys,preprintnumbers,floatfix,nofootinbib,superscriptaddress]{revtex4-2}

\usepackage{amsfonts} % AMS
\usepackage{amssymb} % AMS
\usepackage{amsmath} % AMS
\usepackage{amsthm} % theorems

\usepackage{graphicx} % Include figure files

\usepackage{subcaption}
\usepackage{array} % array
\usepackage{dcolumn} % Align table columns on decimal point
\usepackage{bm} % bold math
\usepackage{latexsym} % latex symbols
\usepackage{hyperref} % hypertext links 
\usepackage{bbold}
\usepackage{color}
\usepackage{multirow}
\usepackage{rotating}
\usepackage{comment}
\usepackage{booktabs}  % for \toprule, \midrule in tables
\usepackage{appendix}

\usepackage[dvipsnames,usenames]{xcolor}
\usepackage{tkz-berge,tikz}
\usepackage[qm,braket]{qcircuit}% quantum circuits

\DeclareMathOperator{\Tr}{Tr}

\definecolor{green}{rgb}{0.1, 0.8, 0.1}

\definecolor{vermillion}{rgb}{0.86, 0.18, 0.01}

\begin{document}

%%%

%\title{Quantum SVM blah blah}
\title{%A Quantum of Cosmos: 
%Quantum Kernel Methods for the Classification of High-dimensional Data on a Superconducting Processor
%
Machine learning of high dimensional data on a noisy quantum processor
}
%
% AUTHOR ORDER - Evan should be up front, then alphabetical, then Gabe last
%
% don't need everyone's email on a long paper - assume Evan is corresponding author...
\author{Evan~Peters}
\email{e6peters@uwaterloo.ca}
\affiliation{Institute for Quantum Computing, University of Waterloo, Waterloo, Ontario, N2L 3G1, Canada}
\affiliation{Department of Applied Mathematics, University of Waterloo, Waterloo, Ontario, N2L 3G1, Canada }
\affiliation{Fermi National Accelerator Laboratory, Batavia, IL 60510}

%\author{Stavros~Efthymiou}
%\email{sefthymiou@google.com}
%\affiliation{X, etc.}

\author{Jo\~ao~Caldeira}
%\email{jcaldeira@gmail.com}
\affiliation{Fermi National Accelerator Laboratory, Batavia, IL 60510}

\author{Alan Ho}
%\email{alankho@google.com}
\affiliation{Google Quantum AI, Venice, CA 90291, United States}

\author{Stefan~Leichenauer}
%\email{sleichenauer@google.com}
\affiliation{Sandbox@Alphabet, Mountain View, CA 94043, United States}

\author{Masoud Mohseni}
%\email{mohseni@google.com}
\affiliation{Google Quantum AI, Venice, CA 90291, United States}

\author{Hartmut Neven}
%\email{neven@google.com}
\affiliation{Google Quantum AI, Venice, CA 90291, United States}

\author{Panagiotis~Spentzouris}
%\email{spentz@fnal.gov}
\affiliation{Fermi National Accelerator Laboratory, Batavia, IL 60510}

\author{Doug Strain}
%\email{dstrain@google.com}
\affiliation{Google Quantum AI, Venice, CA 90291, United States}

\author{Gabriel~N.~Perdue}
%\email{perdue@fnal.gov}
\affiliation{Fermi National Accelerator Laboratory, Batavia, IL 60510}

\preprint{FERMILAB-PUB-20-624-QIS}
% TODO - only bother with PACS numbers if we need them
% \pacs{\FIXME{get PACS numbers}}
%
\keywords{quantum computing, machine learning, kernel methods}
\date{\today}
\begin{abstract}
We present a quantum kernel method for high-dimensional data analysis using Google's universal quantum processor, Sycamore. %Rainbow-23.
This method is successfully applied to the cosmological benchmark of supernova classification using real spectral features with no dimensionality reduction and without vanishing kernel elements.
% GNP - consider this sentence swap:
Instead of using a synthetic dataset of low dimension or pre-processing the data with a classical machine learning algorithm to reduce the data dimension, this experiment demonstrates that machine learning with real, high dimensional data is possible using a quantum processor; but it requires careful attention to shot statistics and mean kernel element size when constructing a circuit ansatz.
%Instead of using the more commonly used synthetic datasets of low dimension, this experiment demonstrates that the analysis of real, high dimensional data requires careful attention to shot statistics and mean kernel element size when constructing a circuit ansatz.
Our experiment utilizes 17 qubits to classify 67 dimensional data - significantly higher dimensionality than the largest prior quantum kernel experiments - resulting in classification accuracy that is competitive with noiseless simulation and comparable classical techniques.

% \masoud{I moved this sentence}
%This work is supported by simulation using a custom noise model for decoherence and readout errors. 
%We show that analysis of this noise model makes it possible to classically mitigate noise and interpret quantum computational results for kernel methods even without quantum error correction.
%We explore the potential for quantum kernels in data analysis using scientific data without dimensionality reduction and achieve results competitive with classical algorithms.
\end{abstract}

\maketitle

\section{Introduction}

Quantum kernel methods (QKM) \cite{Havlicek2019,PhysRevLett.122.040504} provide techniques for utilizing a quantum co-processor in a machine learning setting.
These methods were recently proven to provide a speedup over classical methods for certain specific input data classes \cite{liu2020rigorous}.
They have also been used to quantify the computational power of data in quantum machine learning algorithms and drive the conditions under which quantum models will be capable of outperforming classical ones \cite{huang2020power}. 
Prior experimental work \cite{kusumoto2019experimental,bartkiewicz2020experimental,Havlicek2019} has focused on artificial or heavily pre-processed data, hardware implementations involving very few qubits, or circuit connectivity unsuitable for NISQ \cite{Preskill2018quantumcomputingin} processors; recent experimental results show potential for many-qubit applications of QKM to high energy physics \cite{wu2020application}.

In this work, we extend the method of machine learning based on quantum kernel methods up to 17 hardware qubits requiring only nearest-neighbor connectivity.
We use this circuit structure to prepare a kernel matrix for a classical support vector machine to learn patterns in 67-dimensional supernova data for which competitive classical classifiers fail to achieve 100\% accuracy.
To extract useful information from a processor without quantum error correction (QEC), we implement error mitigation techniques specific to the QKM algorithm and experimentally demonstrate the algorithm's robustness to some of the device noise. Additionally, we justify our circuit design based on its ability to produce large kernel magnitudes that can be sampled to high statistical certainty with relatively short experimental runs.

We implement this algorithm on the Google Sycamore processor which we accessed through Google's Quantum Computing Service.
This machine is similar to the quantum supremacy demonstration Sycamore chip \cite{Arute2019}, but with only 23 qubits active.
We achieve competitive results on a nontrivial classical dataset, and find intriguing classifier robustness in the face of moderate circuit fidelity.
Our results motivate further theoretical work on noisy kernel methods and on techniques for operating on real, high-dimensional data without additional classical pre-processing or dimensionality reduction.

\section{Quantum kernel Support Vector Machines} 

A common task in machine learning is \textit{supervised learning}, wherein an algorithm consumes datum-label pairs $(x, y) \in \mathcal{X} \times \{0, 1\}$ and outputs a function $f: \mathcal{X} \rightarrow \{0, 1\}$ that ideally predicts labels for seen (training) input data and generalizes well to unseen (test) data.
A popular supervised learning algorithm is the Support Vector Machine (SVM) \cite{cortes1995support,Boser:1992:TAO:130385.130401} which is trained on inner products $\langle x_i, x_j\rangle$ in the input space to find a robust linear classification boundary that best separates the data. An important technique for generalizing SVM classifiers to non-linearly separable data is the so-called ``kernel trick''  which replaces  $\langle x_i, x_j\rangle$ in the SVM formulation by a symmetric positive definite kernel function $k(x_i, x_j)$ \cite{Aizerman1964}. Since every kernel function corresponds to an inner product on input data mapped into a feature Hilbert space \cite{aronszajn1950theory}, linear classification boundaries found by an SVM trained on a high-dimensional mapping correspond to complex, non-linear functions in the input space. 
\onecolumngrid

\begin{figure}[t]
    \centering
    \includegraphics[width=0.91\textwidth]{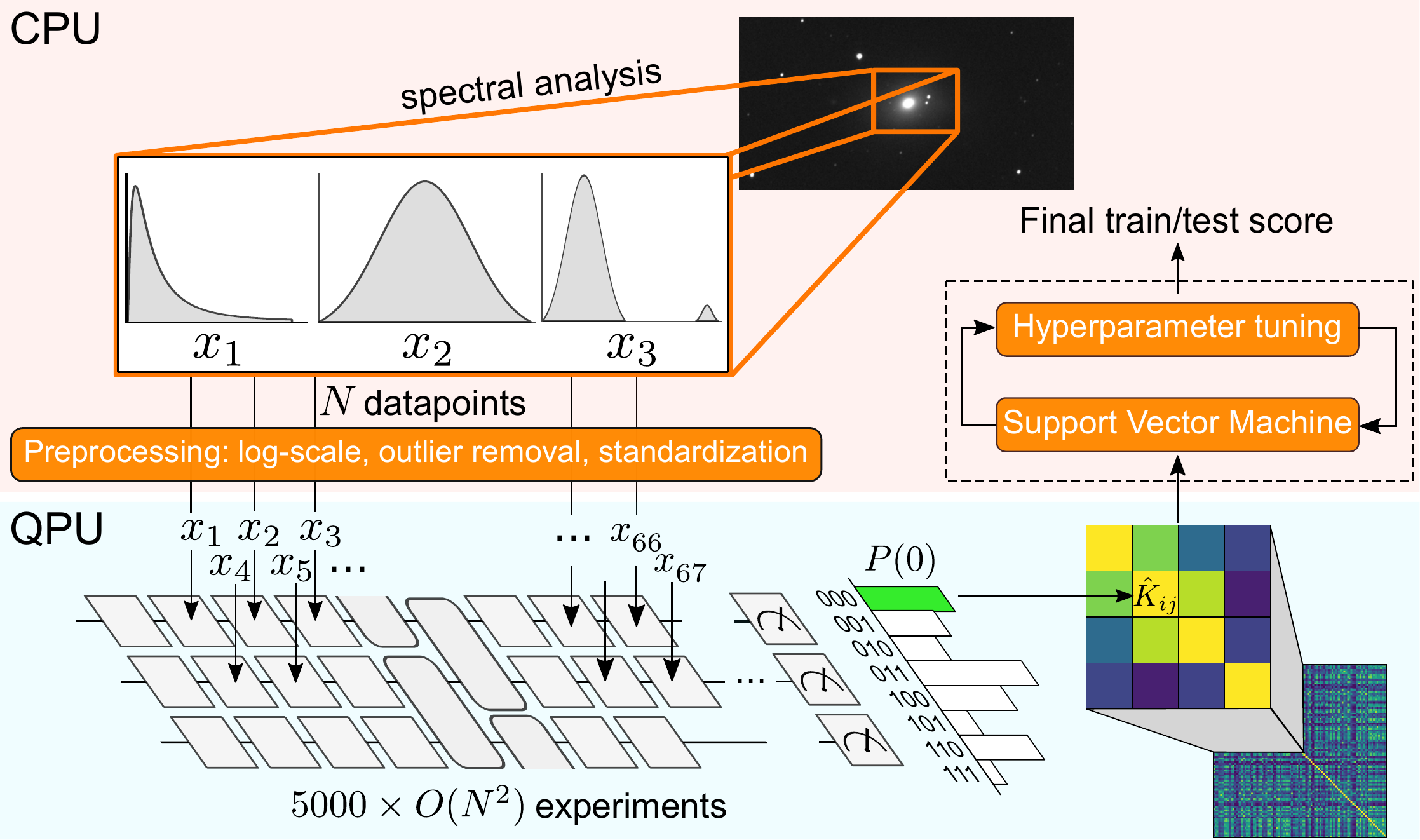}
    \caption{In this experiment we performed limited data preprocessing that is standard for state-of-the-art classical techniques, before using the quantum processor to estimate the kernel matrix $\hat{K}_{ij}$ for all pairs of encoded datapoints $(x_i, x_j)$ in each dataset. We then passed the kernel matrix back to a classical computer to optimize an SVM using cross validation and hyperparameter tuning before evaluating the SVM to produce a final train/test score.}
    \label{fig:svm_flowchart}

\end{figure}

\twocolumngrid
Quantum kernel methods can potentially improve the performance of classifiers by using a quantum computer to map input data in $\mathcal{X}\subset \mathbb{R}^d$ into a high-dimensional complex Hilbert space, potentially resulting in a kernel function that is expressive and  challenging to compute classically. It is difficult to know without sophisticated knowledge of the data generation process whether a given kernel is particularly suited to a dataset, but perhaps families of classically hard kernels may be shown empirically to offer performance improvements.
In this work we focus on a non-variational quantum kernel method, which uses a quantum circuit $U(x)$ to map real data into quantum state space according to a map $\phi(x) = U (x) |0\rangle$. The  kernel function we employ is then the squared inner product between pairs of mapped input data given by  $k(x_i, x_j) = |\langle \phi(x_i) | \phi(x_j) \rangle|^2$, which allows for more expressive models compared to the alternative choice $\langle \phi (x_i) | \phi (x_j) \rangle$ \cite{huang2020power}.

In the absence of noise, the kernel matrix $K_{ij} = k(x_i, x_j)$ for a fixed dataset can therefore be estimated up to statistical error by using a quantum computer to sample outputs of the circuit $U^\dagger (x_i) U (x_j)$ and then computing the empirical probability of the all-zeros bitstring. However in practice, the kernel matrix $\hat{K}_{ij}$ sampled from the quantum computer may be significantly different from $K_{ij}$ due to device noise and readout error. Once $\hat{K}_{ij}$ is computed for all pairs of input data in the training set, a classical SVM can be trained on the outputs of the quantum computer. An SVM trained on a size-$m$ training set $\mathcal{T} \subset \mathcal{X}$ learns to predict the class $f(x) = \hat{y}$ of an input data point $x$ according to the decision function:
\begin{equation}\label{eq:decision_main}
f(x) = \text{sign}\left(\sum_{i=1}^m \alpha_{i} y_i k(x_i, x) + b\right)
\end{equation}
where $\alpha_i$ and $b$ are parameters determined during the training stage of the SVM. Training and evaluating the SVM on $\mathcal{T}$ requires an $m \times m$ kernel matrix, after which each data point $z$ in the testing set $\mathcal{V}\subset \mathcal{X}$ may be classified using an additional $m$ evaluations of $k(x_i, z)$ for $i=1\dots m$. Figure \ref{fig:svm_flowchart} provides a schematic representation of the process used to train an SVM using quantum kernels. 

% % % Preprocessing data
\subsection{Data and preprocessing} \label{sec:dataset}

We used the dataset provided in the Photometric LSST Astronomical Time-series Classification Challenge (PLAsTiCC) \cite{team2018photometric} that simulates observations of the Vera C. Rubin Observatory \cite{verarubin}. The PLAsTiCC data consists of simulated astronomical time series for several different classes of astronomical objects.
The time series consist of measurements of flux at six wavelength bands.
Here we work on data from the training set of the challenge.
To transform the problem into a binary classification problem, we focus on the two most represented classes, 42 and 90, which correspond to types II and Ia supernovae, respectively.

Each time series can have a different number of flux measurements in each of the six wavelength bands.
In order to classify different time series using an algorithm with a fixed number of inputs, we transform each time series into the same set of derived quantities.
These include: the number of measurements; the minimum, maximum, mean, median, standard deviation, and skew of both flux and flux error; the sum and skew of the ratio between flux and flux error, and of the flux times squared flux ratio; the mean and maximum time between measurements; spectroscopic and photometric redshifts for the host galaxy; the position of each object in the sky; and the first two Fourier coefficients for each band, as well as kurtosis and skewness.
In total, this transformation yields a 67-dimensional vector for each object.

To prepare data for the quantum circuit, we convert lognormal-distributed spectral inputs to $\log$ scale, and normalize all inputs to $\left[-\frac{\pi}{2}, \frac{\pi}{2}\right]$.
We perform no dimensionality reduction.
Our data processing pipeline is consistent with the treatment applied to state-of-the-art classical methods.
Our classical benchmark is a competitive solution to this problem, although significant additional feature engineering leveraging astrophysics domain knowledge could possibly raise the benchmark score by a few percent.

% % % Designing the circuit
\subsection{Circuit design}

\begin{figure}[htbp!]
    \centering
    \includegraphics[width=\columnwidth]{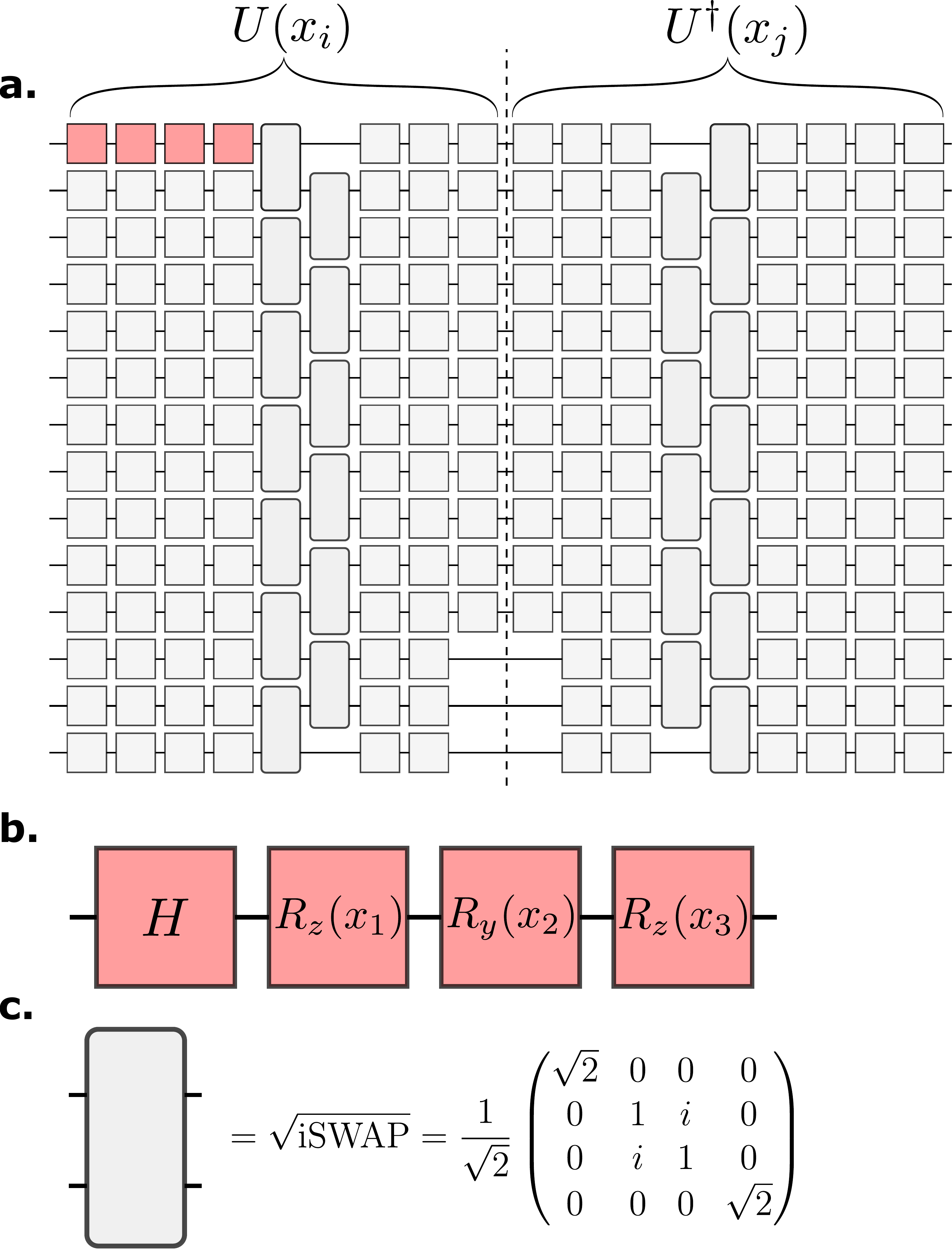}
    \caption{\textbf{a.} 14-qubit example of the type 2 circuit used for experiments in this work. The dashed box indicates $U(x_i)$, while the remainder of the circuit computes $U^\dagger(x_j)$ to ouput $|\langle \phi(x_j)|\phi(x_i)\rangle |^2$. Non-virtual gates occurring at the boundary (dashed line) are contracted for hardware runs. \textbf{b.} The basic encoding block consists of a Hadamard followed by three single-qubit rotations, each parameterized by a different element of the input data $x$ (normalization and encoding constants omitted here). \textbf{c.} We used the $\sqrt{\text{iSWAP}}$ entangling gate, a hardware-native two-qubit gate on the Sycamore processor.}
    \label{fig:circuit_main}

\end{figure}

To compute the kernel matrix $K_{ij} \equiv k(x_i, x_j)$ over the fixed dataset we must run $R$ repetitions of each circuit $U^\dagger (x_j) U(x_i)$ to determine the total counts $\nu_0$ of the all zeros bitstring, resulting in an estimator $\hat{K}_{ij} = \frac{\nu_0}{R}$. This introduces a challenge since quantum kernels must also be sampled from hardware with low enough statistical uncertainty to recover a classifier with similar performance to noiseless conditions. Since the likelihood of large relative statistical error between $K$ and $\hat{K}$ grows with decreasing magnitude of $\hat{K}$ and decreasing $R$, the performance of the hardware-based classifier will degrade when the kernel matrix to be sampled is populated by small entries. Conversely, large kernel magnitudes are a desirable feature for a successful quantum kernel classifier, and a key goal in circuit design is to balance the requirement of large kernel matrix elements with a choice of mapping that is difficult to compute classically. Another significant design challenge is to construct a circuit that separates data according to class without mapping data so far apart as to lose information about class relationships - an effect sometimes referred to as the ``curse of dimensionality'' in classical machine learning. 

For this experiment, we accounted for these design challenges and the need to accommodate high-dimensional data by mapping data into quantum state space using the quantum circuit shown in Figure \ref{fig:circuit_main}. Each local rotation in the circuit is parameterized by a single element of preprocessed input data so that inner products in the quantum state space correspond to a similarity measure for features in the input space. Importantly, the circuit structure is constrained by matching the input data dimensionality to the number of local rotations so that the circuit depth and qubit count individually do not significantly impact the performance of the SVM classifier in a noiseless setting. This circuit structure consistently results in large magnitude inner products (median $K \geq 10^{\text{-}1}$) resulting in estimates for $\hat{K}$ with very little statistical error. We provide further empirical evidence justifying our choice of circuit in Appendix \ref{app:circuit}.

%% Hardware and optimizations
\section{Hardware classification results}

\subsection{Dataset selection}\label{sec:data_selection}

\begin{figure}
	\centering
	\includegraphics[width=\columnwidth]{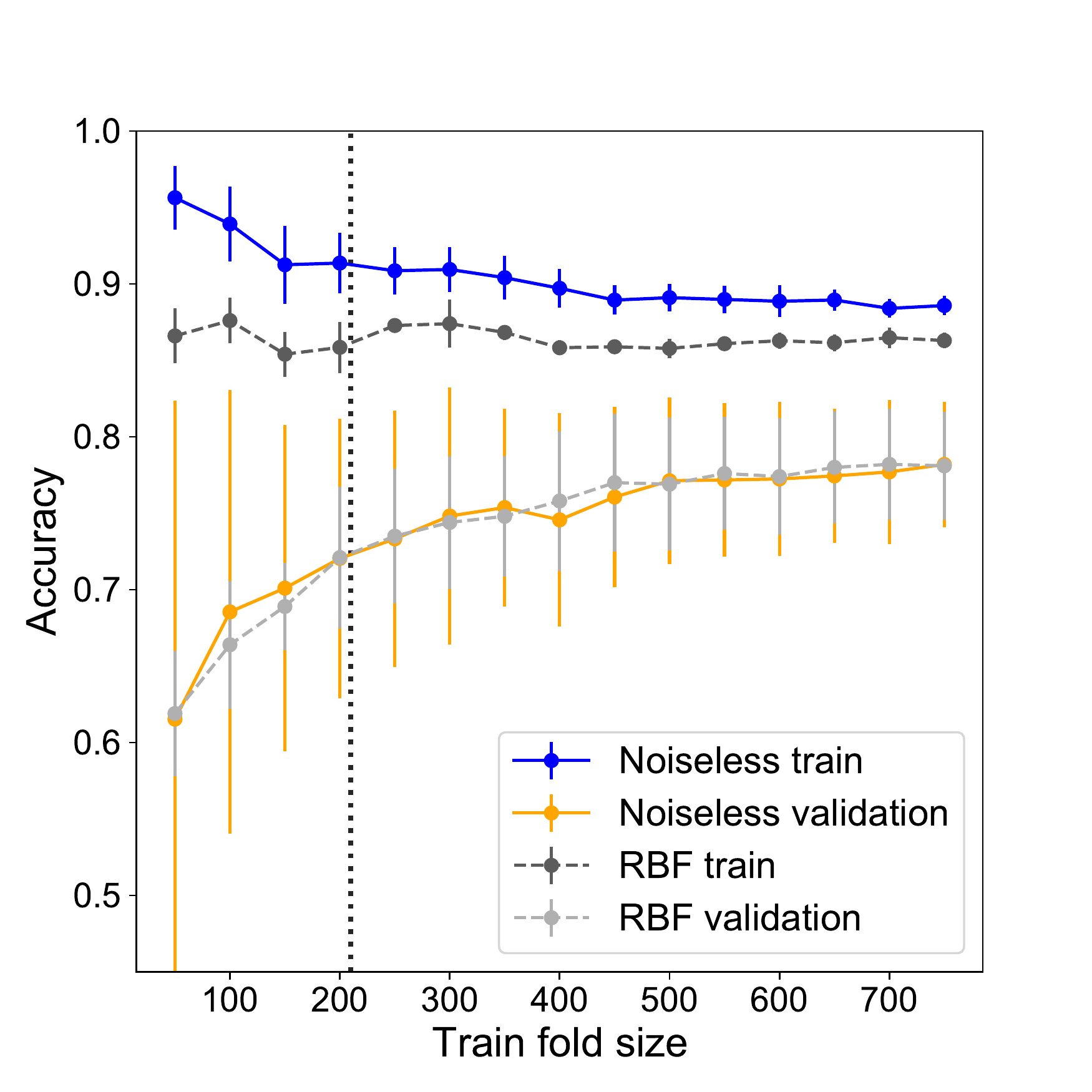}
	\caption{Learning curve for an SVM trained using noiseless circuit encoding on 17 qubits  vs. RBF kernel $k(x_i, x_j) = \exp(-\gamma ||x_i - x_j ||^2)$.  Points reflect train/test accuracy for a classifier trained on a stratified 10-fold split resulting in a size-$x$ balanced subset of preprocessed supernova datapoints. Error bars indicate standard deviation over 10 trials of downsampling, and the dashed line indicates the size $m=210$ of the training set chosen for this experiment.
% DON'T DELETE:
% circuit params: with hyperparameters $c_1=0.25$, $C=4$
% with $\gamma=0.012$, $C=2.0$ optimized over a fine gridsearch on $\gamma \in [10^{\text{-} 5}, 10^{\text{-} 1}]$, $C\in [1, 10^3]$.
	}
	\label{fig:learning_curve}	
\end{figure}

We are motivated to minimize the size  $\mathcal{T}\subset\mathcal{X}$ since the complexity cost of training an SVM on $m$ datapoints scales as $\mathcal{O}(m^2)$. However too small a training sample will result in poor generalization of the trained model, resulting in low quality class predictions for data in the reserved size-$v$ test set $\mathcal{V}$. We explored this tradeoff by simulating the classifiers for varying train set sizes in Cirq \cite{Cirq} to construct learning curves (Figure \ref{fig:learning_curve}) standard in machine learning. We found that our simulated 17-qubit classifier applied to 67-dimensional supernova data was competitive compared to a classical SVM trained using the Radial Basis Function (RBF) kernel on identical data subsets. For hardware runs, we constructed train/test datasets for which the mean train and k-fold validation scores achieved approximately the mean performance over randomly downsampled data subsets, accounting for the SVM hyperparameter optimization. The final dataset for each choice of qubits was constructed by producing a $1000 \times 1000$ simulated kernel matrix , repeatedly performing 4-fold cross validation on a size-280 subset, and then selecting as the train/test set the exact elements from the fold that resulted in an accuracy closest to the mean validation score over all trials and folds.

%% Postprocessing, hyperparameter tuning
\subsection{Hardware classification and Postprocessing}\label{sec:main_svm}

\begin{figure}[!htbp]

    \centering
    \includegraphics[width=\columnwidth]{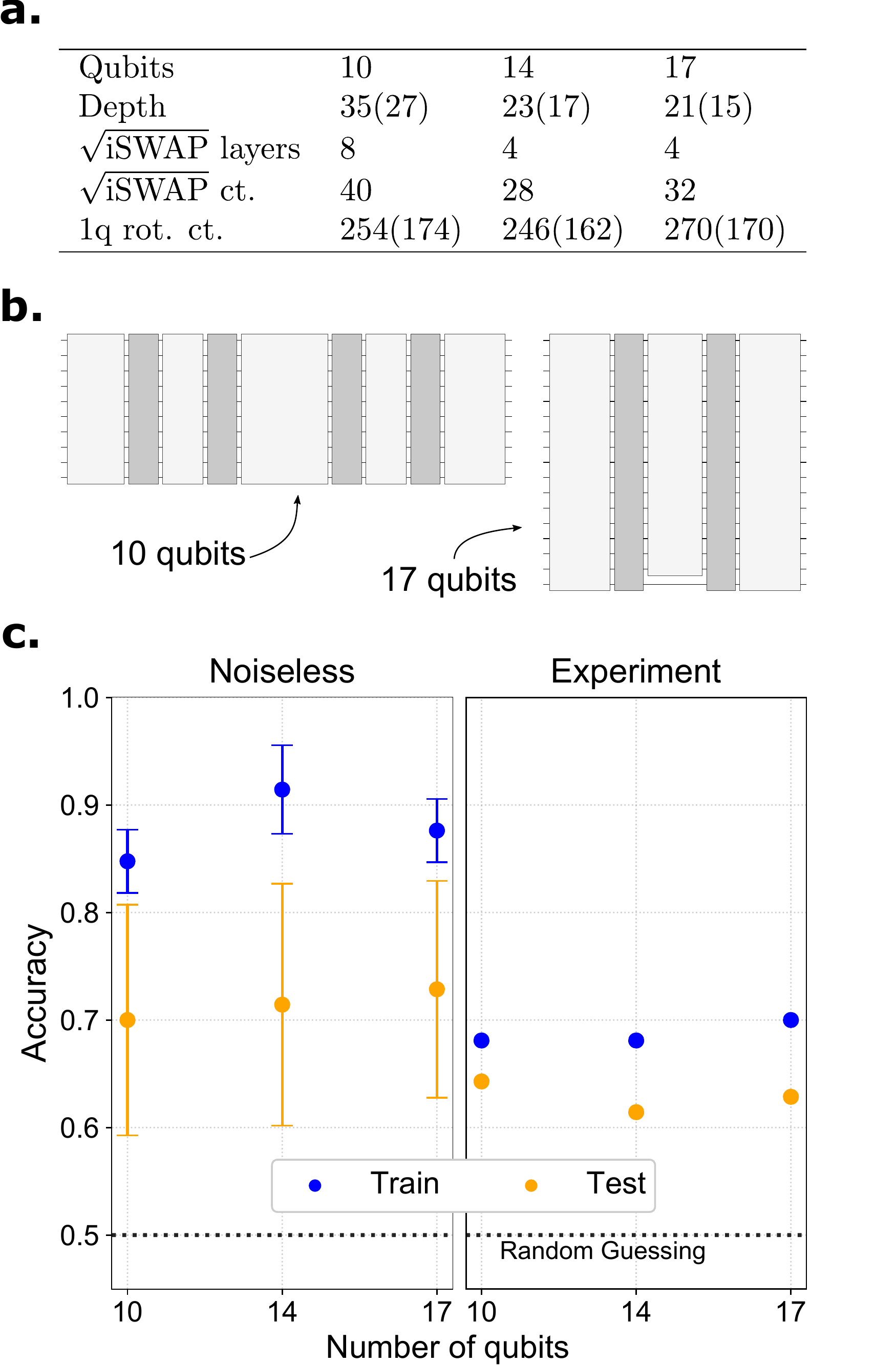}
	\caption{\textbf{a.} Parameters for the three circuits implemented in this experiment. Values in parentheses are calculated ignoring contributions due to virtual Z gates. \textbf{b.} The depth of the each circuit and number of entangling layers (dark grey) scales to accommodate all 67 features of the input data, so that the expressive power of the circuit doesn't change significantly across different numbers of qubits. \textbf{c.} The test accuracy for hardware QKM is competitive with the noiseless simulations even in the case of relatively low circuit fidelity, across multiple choices of qubit counts. The presence of hardware noise significantly reduces the ability of the model to overfit the data. Error bars on simulated data represent standard deviation of accuracy for an ensemble of SVM classifiers trained on 10 size-$m$ downsampled kernel matrices and tested on size-$v$ downsampled test sets (no replacement). Dataset sampling errors are propagated to the hardware outcomes but lack of larger hardware training/test sets prevents appropriate characterization of of a similar margin of error.}
	\label{fig:hero1}	
    % \label{fig:kernel_and_circuits}
\end{figure}

We computed the quantum kernels experimentally using the Google Sycamore processor \cite{Arute2019} accessed through Google's Quantum Computing Service. At the time of experiments, the device consisted of 23 superconducting qubits with nearest neighbor (grid) connectivity. The processor supports single-qubit Pauli gates with $>99\%$ randomized benchmarking fidelity and $\sqrt{i\text{SWAP}}$ native entangling gates with XEB fidelities \cite{Neill195,Arute2019} typically greater than $97\%$.

To test our classifier performance on hardware, we trained a quantum kernel SVM using $n$ qubit circuits for $n\in\{10, 14, 17\}$ on $d=67$ supernova data with balanced class priors using a $m=210, v=70$ train/test split. We ran 5000 repetitions per circuit for a total of $m(m-1)/2 + mv \approx 1.83 \times 10^8$ experiments per number of qubits. As described in Section \ref{sec:data_selection}, the train and test sets were constructed to provide a faithful representation of classifier accuracy applied to datasets of restricted size. Typically the time cost of computing the decision function (Equation \ref{eq:decision_main}) is reduced to some fraction of $mv$ since only a small subset of training inputs are selected as support vectors. However in hardware experiments we observed that a large fraction ($>90 \%$) of data in $\mathcal{T}$ were selected as support vectors, likely due to a combination of a complex decision boundary and noise in the calculation of $\hat{K}$.

Training the SVM classifier in postprocessing required choosing a single hyperparameter $C$ that applies a penalty for misclassification, which can significantly affect the noise robustness of the final classifier. To determine $C$ without overfitting the model, we performed leave-one-out cross validation (LOOCV) on $\mathcal{T}$ to determine $C_{opt}$ corresponding to the maximum mean LOOCV score. We then fixed $C=C_{opt}$ to evaluate the test accuracy $\frac{1}{v}\sum_{j=1}^v \Pr( f(x_j)\neq y_j)$ on reserved datapoints taken from $\mathcal{V}$. Figure \ref{fig:hero1} shows the classifier accuracies for each number of qubits, and demonstrates that the performance of the QKM is not restricted by the number of qubits used. Significantly, the QKM classifier performs reasonably well even when observed bitstring probabilities (and therefore $\hat{K}_{ij}$) are suppressed by a factor of 50\%-70\% due to limited circuit fidelity. This is due in part to the fact that the SVM decision function is invariant under scaling transformations $K \rightarrow r K$ and highlights the noise robustness of quantum kernel methods.

\section{Conclusion and outlook}

Whether and how quantum computing will contribute to machine learning for real world classical datasets remains to be seen. In this work, we have demonstrated that quantum machine learning at an intermediate scale (10 to 17 qubits) can work on “natural” datasets using Google’s superconducting quantum computer. In particular, we presented a novel circuit ansatz capable of processing high-dimensional data from a real-world scientific experiment without dimensionality reduction or significant pre-processing on input data, and without the requirement that the number of qubits matches the data dimensionality. We demonstrated classification results that were competitive with noiseless simulation despite hardware noise and lack of quantum error correction. While the circuits we implemented are not candidates for demonstrating quantum advantage, these findings suggest quantum kernel methods may be capable of achieving high classification accuracy on near-term devices. 

Careful attention must be paid to the impact of shot statistics and kernel element magnitudes when evaluating the performance of quantum kernel methods. This work highlights the need for further theoretical investigation under these constraints, as well as motivates further studies in the properties of noisy kernels. 

The main open problem is to identify a “natural” data set that could lead to beyond-classical performance for quantum machine learning. We believe that this can be achieved on datasets that demonstrate correlations that are inherently difficult to represent or store on a classical computer, hence inherently difficult or inefficient to learn/infer on a classical computer. This could include quantum data from simulations of quantum many-body systems near a critical point or solving linear and nonlinear systems of equations on a quantum computer \cite{Kiani2020, lloyd2020quantum}. The quantum data could be also generated from quantum sensing and quantum communication applications. The software library TensorFlow Quantum (TFQ) \cite{TFQ2020} was recently developed to facilitate the exploration of various combinations of data, models, and algorithms for quantum machine learning. Very recently, a quantum advantage has been proposed for some engineered dataset and numerically validated on up to 30 qubits in TFQ using similar quantum kernel methods as described in this experimental demonstration \cite{huang2020power}. These developments in quantum machine learning alongside the experimental results of this work suggest the exciting possibility for realizing quantum advantage with quantum machine learning on near term processors.

\begin{acknowledgments}
We would like to thank Google Quantum AI team for time on their Sycamore-chip quantum computer. In particular, the presentation and discussion with Kostyantyn Kechedzhi on error mitigation techniques that was incorporated into this experiment and Ping Yeh's participation in some of the group discussions.
Pedram Roushan provided a great deal of useful feedback on early versions of the draft and joined in several useful discussions.
We would also like to thank Stavros Efthymiou for some early work on the quantum circuit simulations, and Brian Nord for consultation on interesting datasets in the domain of astrophysics and cosmology.

EP is partially supported through A Kempf's Google Faculty Award. JC, GP, and EP are partially supported by the DOE/HEP QuantISED program grant HEP Machine Learning and Optimization Go Quantum, identification number 0000240323.
This manuscript has been authored by Fermi Research Alliance, LLC under Contract No. DE-AC02-07CH11359 with the U.S. Department of Energy, Office of Science, Office of High Energy Physics.
\end{acknowledgments}

%
%-----------
% reference
%-----------
\bibliography{quantumbib} %%% ref.bib file

\newpage
\onecolumngrid
\appendix
\begin{center}
\section{Binary classification with Support Vector Machines}
\end{center}

 Supervised learning algorithms are tasked with the following problem: Given input data $\mathcal{X} \subset \mathbb{R}^d$ composed of $d$-dimensional datapoints and the corresponding class labels taken from $\mathcal{Y} = \{-1, 1\}$ attached to each datapoint, construct a function $f$ that can successfully predict $f(x_i) = y_i$ given a datapoint-label pair taken from the dataset $(x_i, y_i) \in \mathcal{X} \times \mathcal{Y}$.

We now introduce the theoretical foundations for the Support Vector Machine, or SVM (see \cite{burges1998tutorial,shawe2004kernel} for a thorough review). A linear SVM performs binary classification by constructing a $(d-1)$ dimensional hyperplane $\langle x, w \rangle + b$ that divides elements of $\mathcal{X}$ according to their class, by finding the hyperplane with the largest perpendicular distance (``margin'') to elements of either class. The hyperplane parameters capable of classifying linearly separable data must satisfy the inequality
\begin{equation}\label{eq:primal_constraint}
    y_i \left(\langle x_i, w \rangle + b \right) \geq 1
\end{equation}
which corresponds to a symmetric margin of $2 / ||w ||$ dividing classes of linearly separable data. Maximizing the margin therefore corresponds to minimizing the hyperplane normal vector, so the task of the SVM is to find the solution to the convex problem

\begin{equation}\label{eq:w}
    \min_{w,b} \frac{1}{2}||w||^2
\end{equation}

Equations \ref{eq:primal_constraint}-\ref{eq:w} frame a constrained optimization problem that can be solved by method of Lagrange multipliers. The Lagrangian to minimize is then
\begin{equation}\label{eq:primal_lagrangian}
    L_P = \frac{1}{2} ||w||^2 - \sum_{i=1} \alpha_i y_i (\langle x_i, w \rangle + b) + \sum_{i=1} \alpha_i
\end{equation}

Recognizing that the inequality of Equation \ref{eq:primal_constraint} can only be satisfied by fully separable data, SVM classifiers trained on real data typically employ so-called ``slack'' variables that loosen the classification constraints and introduce a misclassification penalty to the Lagrangian formulation \cite{cortes1995support}. The constraints for training a linear SVM on non-separable data using slack variables $\xi_i$ then takes on the form:
\begin{align}
    y_i \left( \sum_{s\in SV} \alpha_s y_s k(x_i, x_s) + b \right) &\geq 1 - \xi_i \\
    \xi_i &\geq 0 \quad \forall i
\end{align}

where we choose to assign an L2 penalty for misclassification by adding an additional cost term to the objective function \ref{eq:w}, resulting in the modified objective function

\begin{equation}\label{eq:L2_w}
    \min_{w,b} \frac{1}{2}||w||^2 + \frac{C}{2}\sum_i \xi_i^2
\end{equation}
 
Applying the method of Lagrange multipliers, the primal Lagrangian for Equation \ref{eq:L2_w} is

\begin{equation}\label{eq:L2_primal}
    L_P =\frac{1}{2}||w||^2 - \sum_{i=1} \alpha_i (y_i (\langle x_i, w \rangle + b) - 1 + \xi_i) - \sum_i \mu_i \xi_i + \frac{C}{2}\sum_i \xi_i^2
\end{equation}

Alternatively this can be reformulated to the Wolfe dual problem \cite{Fletcher1987}, with the goal of \textit{maximizing} the dual Lagrangian
% \begin{equation}
%     L_D = \sum_i \alpha_i - \frac{1}{2} \sum_{i,j} \alpha_i \alpha_j y_i y_j \langle x_i, x_j \rangle 
% \end{equation}
% Solving for the dual form of Equation \ref{eq:L2_primal}, training the SVM on a dataset then consists of maximizing the dual Lagrangian:
\begin{equation}\label{eq:L2_dual}
    L_D = \sum_i \alpha_i - \frac{1}{2} \sum_{i,j} \alpha_i \alpha_j y_i y_j \langle x_i, x_j \rangle - \frac{C}{2} \sum_i \xi_i^2
\end{equation}
subject to the constraints:
\begin{align}\label{eq:constraint1}
    0 \leq \alpha_i &\leq C \\\label{eq:constraint2}
    \sum_i \alpha_i y_i &= 0
\end{align}

As Equation \ref{eq:L2_w} is a convex programming problem, the solutions to the primal Lagrangian of Equation \ref{eq:L2_primal} and the dual Lagrangian of Equation \ref{eq:L2_dual} are subject to the Karush-Kuhn-Tucker (KKT) conditions \cite{kuhn1951}:

\begin{align}\label{eq:KKT1}
    w - \sum_i \alpha_i y_i x_i = 0& \\ 
    \sum_i \alpha_i y_i = 0& \\ \label{eq:KKT3}
    C - \alpha_i - \mu_i = 0& \\
    \alpha_i (y_i (\langle x_i, w \rangle + b) - 1 + \xi_i) = 0& \\
    y_i (\langle x_i, w \rangle + b) - 1 + \xi_i \geq 0& \\ \label{eq:KKT6}
    \mu_i \xi_i = 0& \\
    C \geq \alpha_i \geq 0& \\
    \mu_i \geq 0& \\\label{eq:KKT9}
    \xi_i \geq 0&
\end{align}

These conditions determine the hyperplane intercept $b$ and also describe the geometry of the maximal margin hyperplane for the trained SVM. Once the optimal set of parameters $\vec{\alpha}$ is determined with respect to the training inputs $\mathcal{X}$, the linear SVM predicts the class of a data point using the decision function
\begin{equation}\label{eq:decision}
    f(x_p) = \sum_{s \in SV} \alpha_s y_s \langle x_p, x_s\rangle + b
\end{equation}
where the sum runs over the indices of the support vectors, or equivalently all nonzero $\alpha_i$.

 Equation \ref{eq:decision} and the dual Lagrangian \ref{eq:L2_dual} no longer explicitly reference elements of the input space $w, x_i \in \mathbb{R}$, so the optimization problem is still valid under the substitution $x \rightarrow \phi(x)$ for some mapping $\phi: \mathcal{X} \rightarrow \mathcal{H}$ where $\mathcal{H}$ is a Hilbert space (this is the so-called ``kernel trick'' \cite{Aizerman1964}). This permits us to embed input data into higher dimensional Hilbert space for some choice of $\phi$ and then train the SVM on inner products in the mapped space, $\langle \phi(x_i), \phi(x_j) \rangle_\mathcal{H} \equiv k(x_i, x_j)$, where the status of $k$ as an inner product guarantees that it is symmetric, positive-definite. The resulting SVM will then be capable of constructing decision boundaries that are nonlinear in the input space $\mathbb{R}^d$ resulting in a decision function
\begin{equation}\label{eq:decision_kernel}
    f(x_p) = \sum_{s \in SV} \alpha_s y_s k( x_p, x_s) + b
\end{equation}

Evaluating $k(x_p, x_s) = K_{ps}$ for a fixed set $x_p,x_s \in \mathcal{T}\cup \mathcal{V}$ recovers Equation \ref{eq:decision_main} from the main body (since $\alpha_i = 0$ for $i$ outside the support vector set by Equations \ref{eq:KKT3} and \ref{eq:KKT6}).

% % % % % % % % % % % % % % % % % % % % % % % % % % % % % % % % 
\vspace*{25px}
{\centering \section{Circuit structure and Hilbert space embedding}}\label{app:circuit}

 \subsection{Statistical uncertainty and vanishing kernels}

We now discuss limitations to hardware-based quantum kernel methods due to statistical uncertainty. Recall that each kernel matrix element $K_{ij} = k(x_i, x_j)$ is computed by sampling the output of a circuit $U^\dagger (x_j) U(x_i)$ for a total of $R$ repetitions and counting the number $\nu_0$ of all-zeros bitstrings that appear. 
This experiment constitutes $R$ trials of a Bernoulli process parameterized by $K_{ij}$; the unbiased estimators for $K_{ij}$ and associated variance are therefore given by:

\begin{align}
	\hat{K}_{ij} &= \frac{\nu_0}{R} \\ \label{eq:vark}
	\text{Var}(\hat{K}_{ij}) &= \frac{\hat{K}_{ij}(1-\hat{K}_{ij})}{R-1}
\end{align}

Note that positive definiteness of $\hat{K}$ is not necessarily preserved in the presence of statistical error and hardware noise, but in practice we found this had little effect on the ability of the SVM to classify data. The $O(R^{-1/2})$ sampling error of Equation \ref{eq:vark} combined with a requirement that $||\hat{K} - K||_F = \left(\sum_{ij} |K_{ij} - \hat{K}_{ij}|\right)^{1/2} \leq \epsilon m$ (where $m=|\mathcal{X}|$) would suggest that an $\epsilon$-close estimation of $K$ could be achieved using $R = O(\epsilon^{-2}N^2)$ shots per kernel element. In the main body we argue that this fails to bound \textit{relative} error between kernel matrices. This is evident in the symmetric Chernoff bound for the relative error of a sampled $\hat{K}_{ij}$ \cite{goemans2015}:
\begin{equation}\label{eq:kernel_chernoff}
    P\left(\frac{|\hat{K} - K |}{K} \geq \varepsilon \right) \leq 2 e^{-R K \varepsilon^2/3}
\end{equation}
for which the probablity of large relative error quickly becomes unbounded for $R << \mathcal{O}(K_{ij}^{-1})$. Relative error is relevant by the following reasoning: Let  $L_D '$ be the L1-penalized dual Lagrangian corresponding to a kernel constructed from the transformation $K' = rK$, given by

\begin{equation}\label{eq:ld_prime}
    L_D ' = \sum_i \alpha_i' - \frac{1}{2} \sum_{i,j} \alpha_i' \alpha_j' y_i y_j K_{ij}'
\end{equation}

If $\alpha_{opt}$ contains the parameters maximizing the Lagrangian
\begin{equation}\label{eq:L1_dual}
    L_D = \sum_i \alpha_i - \frac{1}{2} \sum_{i,j} \alpha_i \alpha_j y_i y_j K_{ij}
\end{equation}

then $\alpha_{opt}$ also maximizes the Lagrangian $\frac{1}{r} L_D$ which may be rewritten as follows:

\begin{align}\label{eq:svm_opt}
   \frac{1}{r} L_D = \sum_i \frac{\alpha_i}{r} - \frac{1}{2} \sum_{i,j} \frac{\alpha_i \alpha_j}{r^2} y_i y_j (rK_{ij}) 
\end{align}

By comparison with Equation \ref{eq:ld_prime} $L_D '$ has the immediate choice of unique solution $\alpha_{opt}' = \alpha_{opt} / r$, which may be achieved by appropriate choice of penalty parameter $C$ appearing in the KKT conditions \ref{eq:KKT1}-\ref{eq:KKT9} for $L_D '$ (or equivalently in the primal problem $L_P '$ before kernelizing the problem). A similar result can be attained if an L2 misclassification penalty is used by restricting the Lagrange multipliers $\alpha$. Consequently the decision function for an SVM classifier trained on the kernel matrix $K'$ is identical to the decision function for the SVM classifier trained on the original $K$. This result makes intuitive sense for the choice of a linear kernel $K_{ij} = \langle x_i , x_j \rangle$, for which stretching/shrinking each datapoint $x_i \rightarrow x_i/\sqrt{r}$ has no effect on the geometry of the maximal margin hyperplane. By choosing some $r$ for the transformation $K\rightarrow rK$, $\hat{K}\rightarrow r\hat{K}$, the absolute error $||K - \hat{K}||_F$ may be made arbitrarily large or small without affecting the performance of the associated SVM classifier, which suggests that $||K - \hat{K}||_F$ does not completely characterize the resulting SVM accuracy.

The high dimensionality of quantum state space poses a threat to the experimental feasibility of quantum kernel methods since the relative statistical error incurred by finite shot statistics grows as the magnitude of the sampled kernel element shrinks. Using a naive example,  a randomly selected encoding unitary will almost certainly result in vanishing kernel elements by strong measure concentration: If we treat $S = \{U^\dagger (x_j) U(x_i)\}$ as a distribution of unitaries that are random with respect to the Haar measure it is well known that the expected probability for any specific bitstring (and therefore $\frac{\nu_0}{R}$) sampled from a unitary in $S$ scales as $\mathcal{O}(2^{-n})$. In practice, the preprocessing of input data and circuit structure must be chosen with careful attention given to the corresponding distribution on $K$.

\begin{figure}[!htb]
    \centering
    \includegraphics[width=\textwidth]{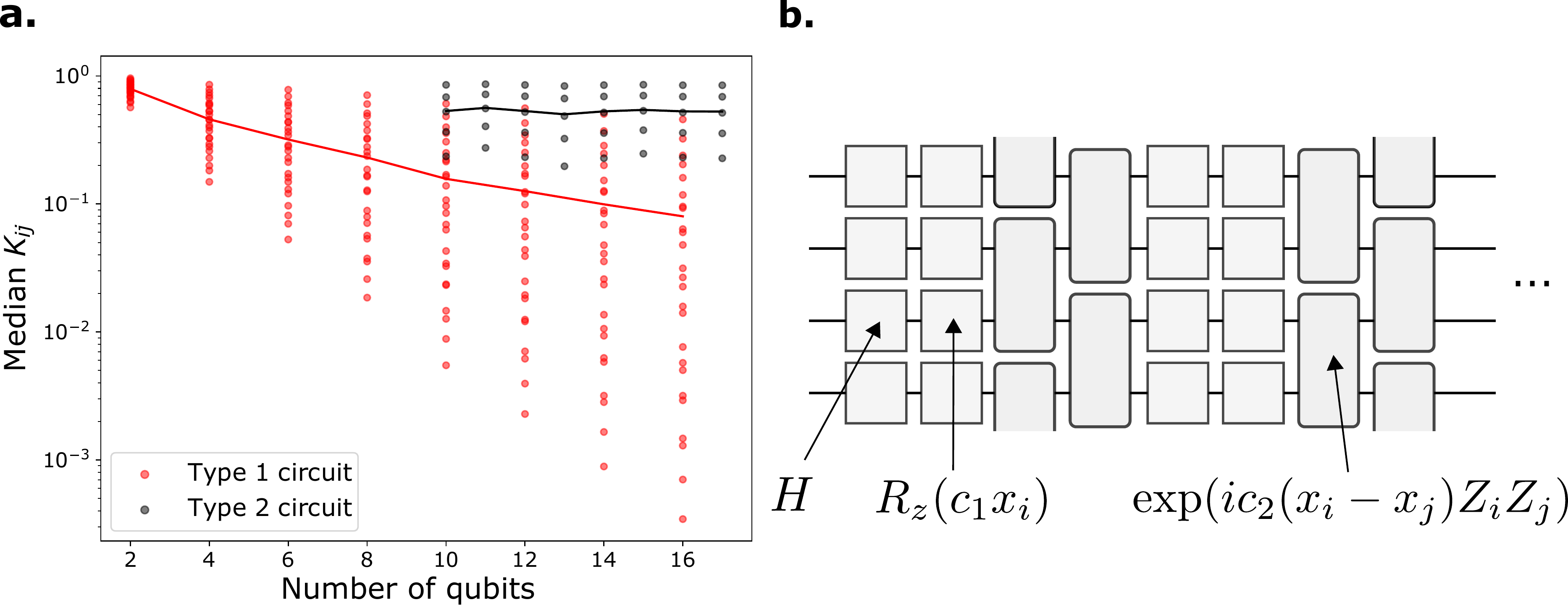}
    \caption{Circuit structure and data preprocessing has a large impact on the resulting distributions of kernel matrix elements. (a) Distributions of median $K$ with respect to a coarse grid search over $c_1, c_2 \in \{0.1, 0.15, 0.2, 0.25, 0.3\}$ for type 1 circuits with $n$-dimensional PCA compressions as input suggest that vanishing kernel magnitudes (red) make much of the gridsearch space inaccesible to realistic hardware experiments for even modest numbers of qubits. We found no such trend in $K$ for type 2 circuits (Equation \ref{fig:circuit_main}) implemented in our experiments with 67-dimensional input data.}
    \label{fig:kernel_and_circuits}
\end{figure}

To explore this effect we constructed a classifier similar to quantum circuits described in \cite{Havlicek2019, farhi1802classification},  depicted in Figure \ref{fig:kernel_and_circuits}b and given by
\begin{align}\label{eq:type1_circuit}
    U(x) &= H^{\otimes n} V(x) H^{\otimes n} V(x)\\ \nonumber
    V(x) &= \exp \left(\sum_{i=1}^n c_1 x_i Z_i + \sum_{(i, j) \in NN} c_2 (x_i - x_j) Z_i Z_j \right) 
\end{align}

where the entangling gates are selected among nearest neighbors on the (simulated) circuit grid. We chose to parameterize entanglers by $c_2 (x_i - x_j)$ instead of quadratic terms proportional to $x_i x_j$ to eliminate concentration of $E[x_i x_j] \rightarrow 0$ that occurs for our choice of normalization. For clarity we refer to the circuit described by Equation \ref{eq:type1_circuit} as ``Type 1''. An increase in the connectivity results in a higher gate/parameter count, resulting in generally smaller sampled kernel elements. This circuit structure requires that input data have dimension equal to the number of qubits. We applied Principal Component Analysis (PCA) \cite{PCA} to reduce the 67-dimensional data to $n$ dimensions and then standardized the data to the interval $[-\pi/2, \pi/2]$. We defined additional hyperparameters $(c_1, c_2)$ that can be tuned to optimize the cross-validated performance of the corresponding SVM and control the resulting distribution of kernel matrix elements. Figure \ref{fig:kernel_and_circuits}a shows that the magnitude of $K$ vanishes with respect to increasing $c_1, c_2$ or number of qubits $n$. This does not necessarily result low accuracies for the associated SVM classifiers, but describes a family of kernels that are infeasible to sample on hardware. It is possible to preserve the magnitude of K if $c_1$ and $c_2$ are scaled down with increasing $n$ but for small enough angles over/under-rotation errors and noise will become dominating factors in hardware outcomes, while the limit $c_2 \rightarrow 0$ results in a circuit that can be simulated trivially.

These results motivate a new approach for encoding data on large numbers of qubits, especially if the input data dimensionality is large. To compute large-magnitude kernels on high dimensional data without dimensionality reduction, we designed a circuit encoding to map input data $x_i, z_i \in X \subset \mathbb{R}^d$ into a subspace of $\mathbb{C}^{2^n}$ using an approximately orthogonal parameterization of $U(2^n)$, the group describing $n$-qubit unitaries. While examples of exactly orthogonal parameterizations of $U(2^n)$ exist, such as Euler angle parameterization of $U(2^n)$ \cite{Tilma2004}, such schemes are generally inefficient to implement on hardware. We approximate such an encoding using circuits structured similarly to the Hardware Efficient Ansatz \cite{Kandala2017} consisting of an initial layer parameterizing $\bigotimes^n U(2)$ interspersed with local entanglers. This circuit structure (referred to here as ``Type 2'') is shown in Figure \ref{fig:circuit_main} of the main body and can be expressed in terms of individual gates as

\begin{align} \label{eq:type2_circuit}
    U(x) &= \prod_{\ell=1}^L U_B U_A(S_\ell(x)) \\ \nonumber
    U_A(z) &= \bigotimes_{i=1}^n H^{(i)} R_z^{(i)} (c_1 z_{i2}) R_y^{(i)} (c_1 z_{i1}) R_z^{(i)}(c_1 z_{i0}) \\ \nonumber
    U_B &= \prod_{ (i, j) \in E(G)} \sqrt{\text{iSWAP}^{(i,j)}}
\end{align}

where $E(G)$ denotes the set of edges composing a length-$n$ simple path  permitted by the Sycamore connectivity, superscript $(i)$ indicates action on qubit $i$, and $S: \mathbb{R}^d \rightarrow \mathbb{R}^{3n}$ denotes selection of a subset of $3n$ elements from the input data to be encoded into a given rotation layer. The specific choice of rotation and entangling gates was influenced by the gate set available on the processor at the time the experiments were conducted, namely $\sqrt{\text{iSWAP}}$ and the Sycamore gate \cite{Arute2019}. Note that the use of Z rotations in $U_A$ reduces the hardware depth of the corresponding circuit by 50\% \cite{FromDoug}. This architecture therefore encodes $d$-dimensional input data in $\mathcal{O}(d/n)$ depth on hardware. The fact that this circuit choice may be made arbitrarily shallow in number of qubits $n$ rules out the possibility of demonstrating quantum advantage, but the experimental and design challenges of implementing this architecture are relevant to QKM in general.

As depicted in Figure \ref{fig:svm_flowchart}, single qubit rotations in each circuit were ``filled'' sequentially from left to right, top to bottom beginning with the first element $x_1$ of the data and ending with $x_{67}$. Exceptions were made to this pattern to more evenly distribute single qubit rotations between entangling layers. We explored randomized filling schemes and found no significant difference to circuit performance nor inner product magnitudes. When the number of qubits is not an integer factor of 67, gaps appeared in at most two layers of the circuit. 

\subsection{Hyperparameter tuning in the quantum circuit}\label{app:hyperparameter}

\begin{figure}[!htb]
    \centering
    \begin{tabular}[t]{ll}
        \begin{subfigure}[t]{0.45\textwidth}
            \centering
            \includegraphics[width=\textwidth]{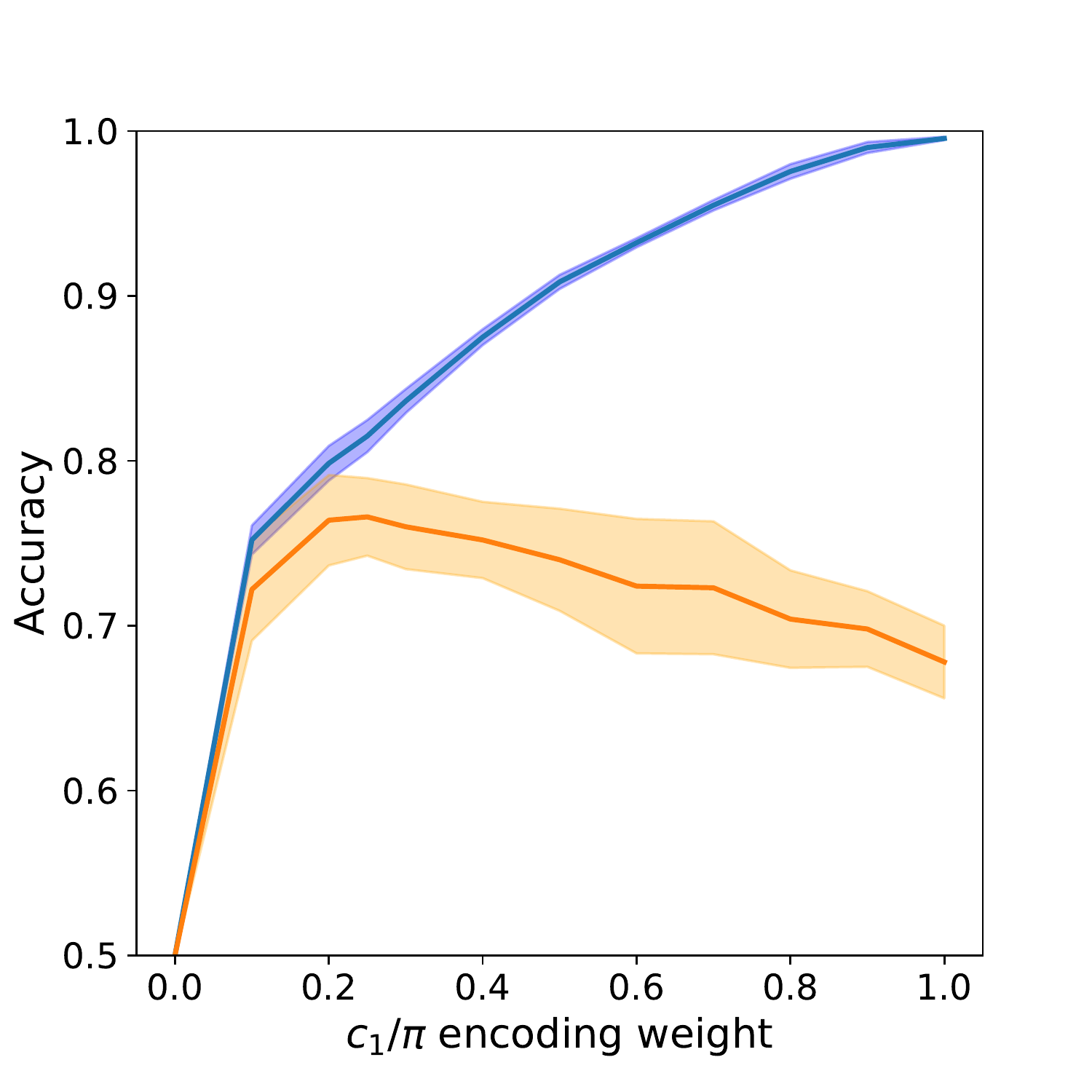}
            % \caption{}
            % \label{fig:circuit2}
        \end{subfigure}
            &
        \begin{subfigure}[t]{0.45\textwidth}
            \centering
            \includegraphics[width=\textwidth]{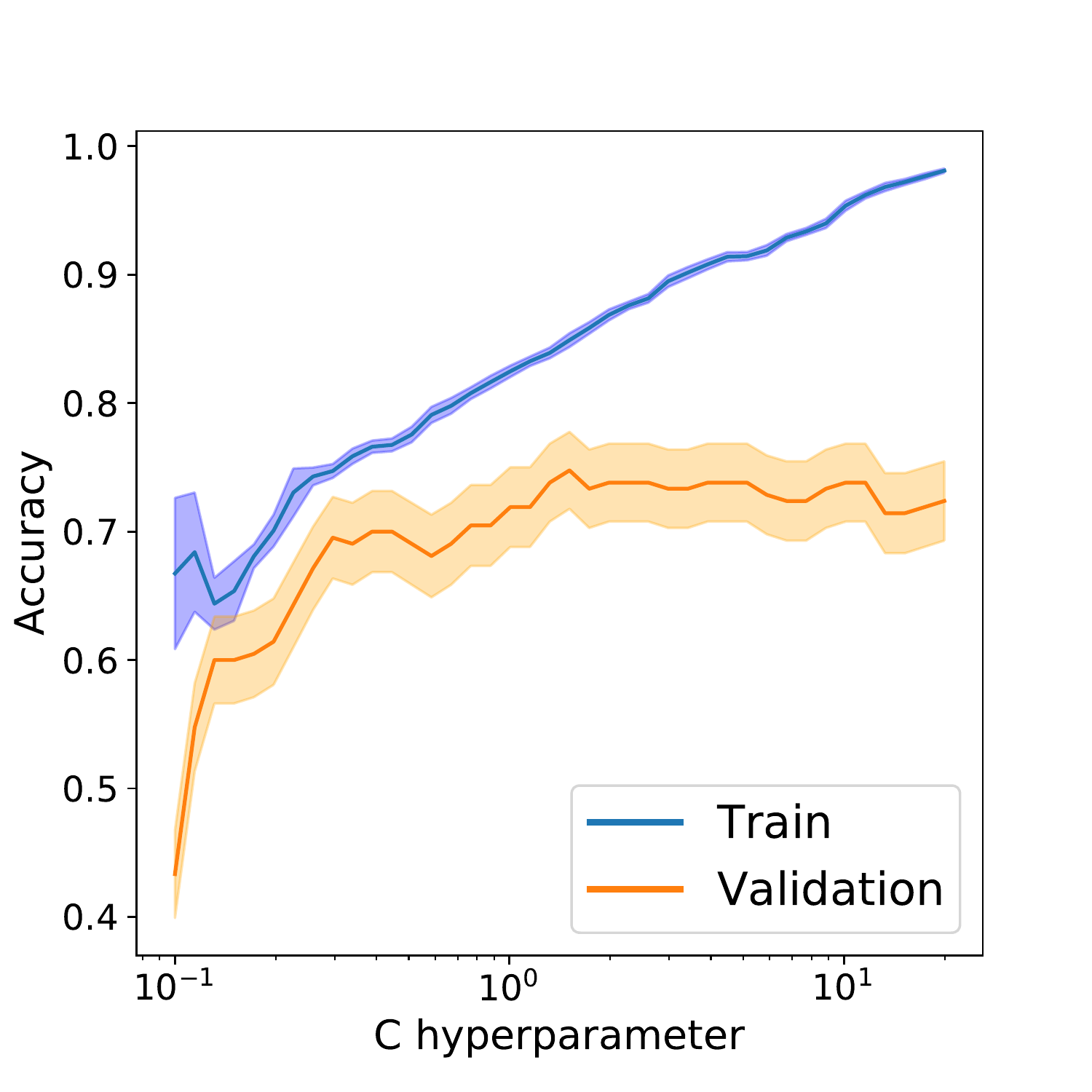}
            % \caption{}
            % \label{fig:circuit1}
        \end{subfigure}
    \end{tabular}
	\caption{Hyperparameter tuning for simulated type 2 circuits consisted of grid search optimization over two parameters. (left) The encoding parameter $c_1$ multiplies each encoded data element and impacts the typical separation (magnitude of $K$) for mapped feature space vectors. (right) The L2 penalty hyperparameter $C$ used to train the SVM allows for robust performance in the presence of noise. Both plots correspond to 10 qubit circuits used in the experiment ($c_1$ was optimized over an $m =1000$ subset of SN-67 dataset while $C$ was optimized over experimental run dataset with $m=210$).}
	\label{fig:sun_c1_tuning}
\end{figure}

Our circuit design introduces a single parameter $c_1$ for multiplicative scaling of on elements of preprocessed $x$. We observed that train/test performance varied with respect to $c_1$ and therefore treated it as a hyperparameter of the kernel function to be optimized in simulation. In practice, the same optimization procedure can be carried out using sweeps over $c_1$ for hardware submissions. Figure \ref{fig:sun_c1_tuning}a shows a typical outcome of the hyperparameter tuning process and demonstrates both a local optimum for validation scores with respect to $c_1$ as well as a capacity for overfitting in the large-$c_1$ limit.

The choice of optimal $c_1$ has a direct impact on the proximity of states $|\psi(x)\rangle$ mapped into the quantum state space; in the trivial limit $c_1\rightarrow 0$ the unitary $U(x)$ becomes the identity map and $K_{ij}\rightarrow 1$; similarly in the large $c_1$ limit the angles between mapped input data grow linearly and $K$ quickly vanishes. Therefore there is a tension between producing mapped states $\{|\psi(x_i)\rangle\}$ with good separability but without the isolation of mapped points in high dimensional space that would degrade performance, a phenomenon in classical machine learning known as the ``curse of dimensionality'' (e.g. \cite{Friedman1994}).

% % % % % % % % % % % % % % % % % % % % % % % % % % % % % % % % 
% \begin{center}
{\centering \section{Dataset selection and preprocessing}}
% \end{center}

We used the dataset provided in the Photometric LSST Astronomical Time-series Classification Challenge (PLAsTiCC) \cite{team2018photometric}. After engineering the 67-float data with binary labels (Section \ref{sec:dataset}), preprocessing consisted of the following steps:
\begin{enumerate}
    \item \textit{Logscale transformation}: Many of the features were distributed in a lognormal distribution which motivates use of the transformation $x_i \rightarrow \log_{10} (x_i)$. Some information loss resulted from taking the absolute value of median flux, for which approximately $4\%$ of entries in the original dataset were negative. All other absolute value operations resulted in negligible loss of sign information. 
    \item \textit{Normalization/scaling and outliers}: Since the local rotations of $U(x)$ are $2\pi$-periodic, an effective normalization scheme is to map the boundaries of the input range to $\left[ -\frac{\pi}{2}, \frac{\pi}{2} \right]$. Realistically, outliers will result in the effective range for most data being compressed into a much smaller space (thereby amplifying the effect of over/under-rotation errors) and so we scaled data in such a way to ignore the effects of large outliers by applying the transformation:
    $$
    x' =  \pi \left(\frac{x - P_1}{P_{99} - P_1}\right) - \frac{\pi}{2}
    $$
    to every element of input data, where $P_k$ denotes the value of the $k$-the percentile of the input domain. This is equivalent to typical implementations of a robust scaler with the quantile range set to (0.01, 0.99) (e.g. \cite{scikit-learn}), which removes the effects of outliers on the rescaling. Then hyperparameter tuning was used to adjust the rotation parameters by a further multiplicative factor $c_1$ (see Appendix \ref{app:hyperparameter}).
\end{enumerate}

% % % % % % % % % % % % % % % % % % % % % % % % % % % % % % % % 
% \begin{center}
{\centering \section{Error mitigation}}
% \end{center}

\subsection{Device parameters}\label{sec:calibration}

Periodic calibrations of the Sycamore superconducting qubit device produce diagnostic data describing qubit and gate performances. Calibration metrics relevant to this experiment included readout errors $p_{00}$ (probability of a computational basis measurement reporting a ``1'' when the result should have been ``0'') and $p_{11}$ (probability of a measurement reporting a ``0'' when the result should have been ``1''), single qubit $T_1$, single qubit gate RB error, and $\sqrt{\text{iSWAP}}$ gate cross-entropy benchmarking (XEB) error. The readout error probabilities were used primarily for constructing and analyzing post-processing techniques for error correction while the remainder of the calibration data were fed in to an automated qubit selection algorithm. 

\subsection{Automated qubit selection}\label{sec:auto_qubit}

\begin{figure}
	\centering
	\includegraphics[width=.45\textwidth]{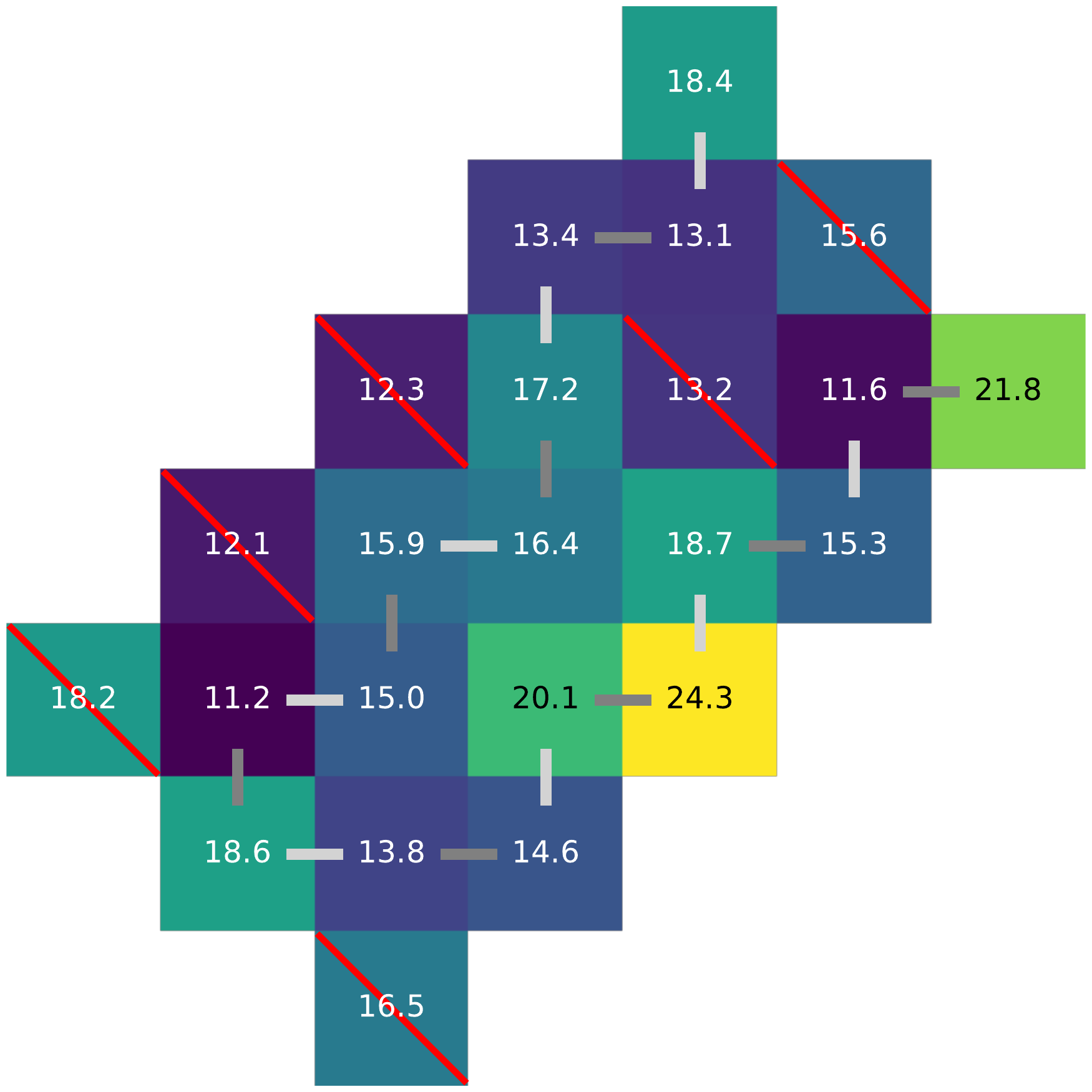}
	\caption{Sample results of automated qubit selection with rejected qubits denoted by a red slash. Entangler patterns (light/dark gray) are overlaid on the Sycamore 23-qubit grid annotated with $T_1$ in $\mu$s. No more than 19 qubits may be assigned to the grid using our connectivity scheme, so that qubit selection has diminishing effects on performance as $n\rightarrow 19$.}
	\label{fig:hardware_grid}	
\end{figure}

To improve the performance of the algorithm for a given number of qubits, we designed a graph traversal algorithm to select qubits based on diagnostic data taken during device calibration. Less weight was applied to $p_0$ and $p_1$ due to the availability of readout error mitigation techniques. We constructed a qubit graph $G_q = (E, V)$ where edges represent entangling gate connectivity and nodes represent qubits according to the Sycamore 23-qubit grid layout. The optimization was done by traversing all simple paths of fixed length $k$ (implemented according to \cite{Sedgewick2001, networkx}), and then scoring each path according to some function of the metrics for the subset of nodes and edges visited. Stated as an optimization problem, given an objective function $f: V \times E \rightarrow \mathbb{R}$ scoring subsets of vertices and edges composing an Eulerian graph, this algorithm finds the maximum evaluation of $f$ over all possible graphs $G_q$:
\begin{equation}
	\max_{f(V_k(G), E_k(G))} G_q
\end{equation}
subject to the constraints $|V_k(G)|=k$, $|E_k(G)| = k-1$. Before applying $f$, the heterogeneous calibration data were normalized to the range $[0, 1]$ and inverted if they represented an error (as opposed to a fidelity). Letting the value of the $p$-th category of calibration data for the i-th qubit $v_i$ be $c_p(v_i)$, and similarly the $p$-th category of calibration for the i-th edge $e_i$ be $d_p(e_i)$, and defining $g_p$ as a scoring function applied to the $p$-th processed calibration metric, our implementation of $f$ takes on the form
\begin{equation}
f(V) = \sum_{p\in C_1} \sum_{v_i \in G}  g_p(c_p(v_i)) + \sum_{p\in C_2} \sum_{e_i \in G}  g_p(c_p(e_i))
\end{equation}

where $C_1$ and $C_2$ represent the calibration metrics corresponding to single and pairs of qubits respectively. We implemented $g_p$ as a logarithmic function for $T_1$, $T_2$, and $f_{XEB,2q}$ metrics and a linear function for $p_{00}$ and $p_{11}$ metrics. Figure \ref{fig:hardware_grid}	 shows the results of an example optimization overlaid on $T_1$ calibration results.

% % % % % % % % % % % % % % % % % % % % % % % % % % % % % % % % 
% Readout error correction
\subsection{Readout error correction}\label{sec:readout_corr}

Readout error resulting from relaxation and thermal excitation can be modelled by a stochastic bitflip process applied to the observed bitstrings. Here we describe an efficient and accurate technique for correcting readout error for quantum kernel methods.

Let $p(y^n|x^n)$ describe the conditional probability for observing bitstring $y^n$ after exposing the bitstring $x^n$ to $n$ distinct bitflip channels, and let $q^k(y|x)$ for $x,y\in\{0,1\}$ describe the corresponding probability for observing bit ``$y$'' after exposing the $k$-th bit ``$x$'' to a single bitflip channel. Then for the $k$-th qubit, the metrics introduced in Appendix \ref{sec:calibration} as $p_{00} = q^k(1|0)$ and $p_{11} = q^k(0|1)$ may be used to partially undo readout error by means of postprocessing. We define a response matrix $R \in \mathbb{R}^{2^n \times 2^n}$ elementwise as $(R)_{xy} = p(y^n|x^n)$ that contains as its elements the total probability for transition from bitstring $x^n=x_1\dots x_n$ to bitstring $y^n=y_1\dots y_n$ computed as the product of $q^k(1|0)$ and $q^k(0|1)$ corresponding to each individual bit:

\begin{equation}\label{eq:bf_prob}
R_{xy} \equiv p(y_1\dots y_n | x_1 \dots x_n) = \prod_{k=1}^n q^k(y_k|x_k)
\end{equation}

For simplicity, we assume that each individual bitflip may be modelled as an independent process, although the techniques discussed here are readily applicable to a system of dependent bitflips if the bitflip likelihoods are experimentally measured in parallel. Then evidently, 
\begin{equation}
    R = \bigotimes_{k=1}^n \begin{pmatrix}
    q^k(0|0) & q^k(0|1) \\ 
    q^k(1|0) & q^k(1|1) 
    \end{pmatrix}
\end{equation}

Note that $R$ is generally asymmetric since typically $q^k(1|0) < q^k(0|1)$. While multiplying $R^{-1}$ by the set of observed bitstring frequencies would recover the prior distribution of bitstring frequencies with good fidelity, standard matrix inversion is subject to instabilities and is not tractable for even modest numbers of qubits. 

Since only the frequency of the all-zero's bitstring is necessary to compute $\hat{K}_{ij}$, we implemented correction using a small subset of bitstring transition probabilities to perform quick and relatively high-fidelity readout error correction in post-processing. We generated $R$ and then truncated the full $2^n$-dimensional basis to the bitstring space containing strings with Hamming weight $\leq k_{max}$ for some $n$-dependent $k_{max}$ resulting in a truncated response matrix $R_{t}$. We then computed the pseudo-inverse $R_t^{-1}$ and performed error correction by simple matrix multiplication on the array of experimental readout frequencies (similarly truncated). This simplification comes at the expense of knowledge about any other post-correction frequencies since the other bitstrings in the truncated space are off-center within the Hamming sphere of kept bitstrings, resulting in a bias in the inverted linear map.

We now analyze the effect of truncation on the readout error correction. The number of simultaneous readout error events (either relaxation or excitation) may be modelled as an induced random variable $Z = \sum_k X_k$ for $\text{Pr}(X_k=1) = q^k(\neg x|x)$. This distribution has expected value $\mu = \sum_k q^k(\neg x|x)$ and an exponentially suppressed likelihood for simultaneous readout errors via the Chernoff bound $\text{Pr}(Z\geq k) \leq \exp(k - \mu - k\log (k/\mu))$. Thus a natural measure for the effect of Hamming weight truncation is the empirical probability allocated to the complement of the truncated subspace:
\begin{equation}\label{eq:trunc_pr}
    \text{Pr}(Z > k_{max}) = 1 - \sum_{i=0}^{k_{max}} \text{Pr}(Z=i)
\end{equation}

\begin{figure}[!htb]
	\centering
	\includegraphics[width=.9\textwidth]{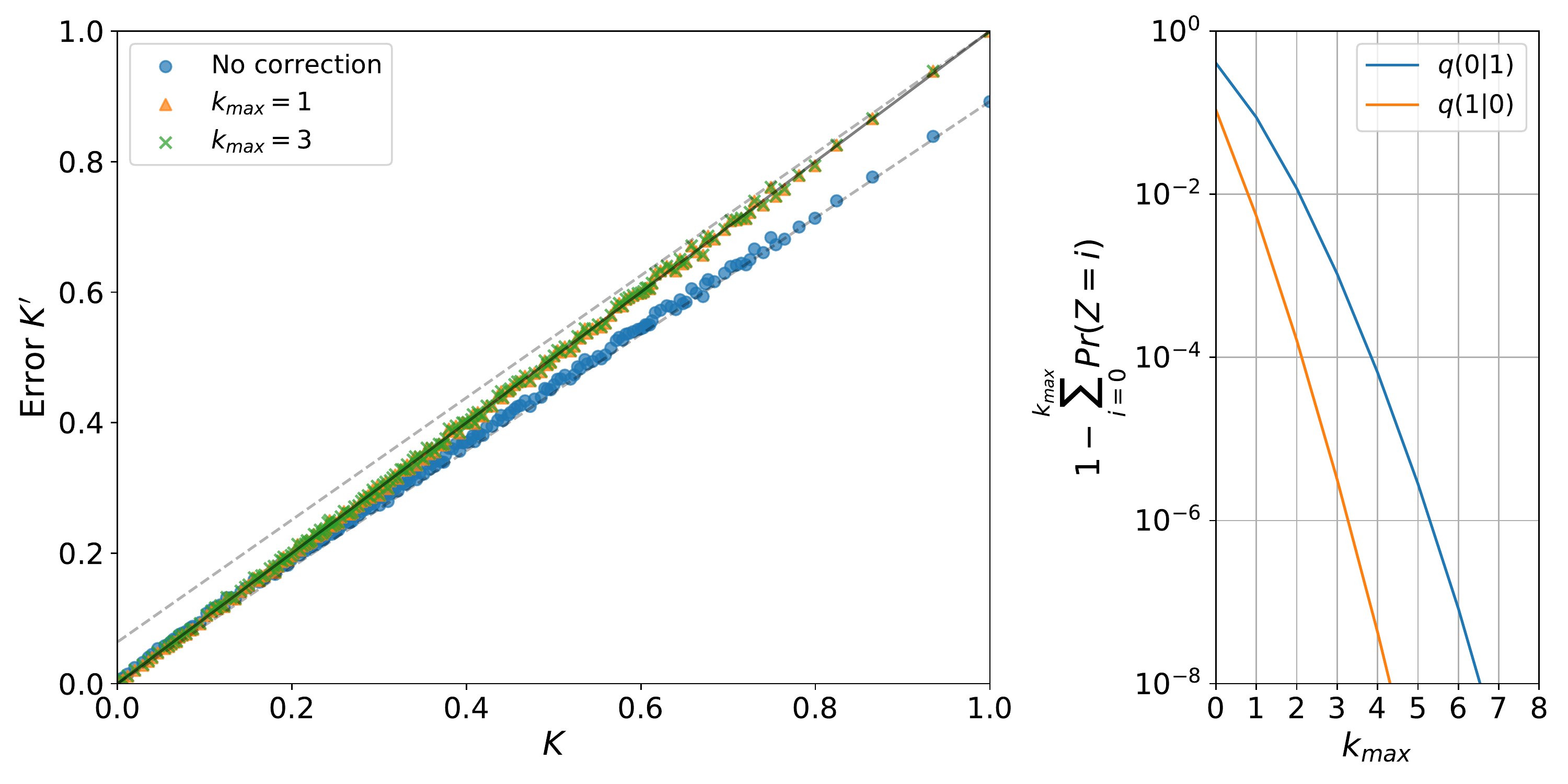}
	\caption{Truncated readout error correction for 10 qubit circuit. (left) Correcting on the likely subspace described by $k_{max}=1$ provides significant error correction, and further increasing $k_{max}$ provides diminishing returns. Dashed lines indicate the infinite-shot upper/lower bounds given by Equation \ref{eq:bf_bounds} (bounds are violated when empirical bitflip probabilities do not match imposed readout error probabilities); black line indicates perfect error correction. (right) Empirical probability for transition out of the weight $\leq k_{max}$ vanishes exponentially in the highest weight considered.}
	\label{fig:bf_sim_ana}	
\end{figure}

While Equation \ref{eq:trunc_pr} describes the probability of events outside the truncated subspace, it does not directly translate to failure probability for truncated readout correction. To explore this effect numerically, we computed the output probability distributions for a 10-qubit quantum kernel circuit sampled for $5000$ repetitions, and then introduced artificial readout error (using bitflip probabilities taken from the Sycamore processor) followed by roughly 5\%  Gaussian noise on the sampled distribution. Figure \ref{fig:bf_sim_ana} shows the error distribution as well as the effect of correcting using an inverted truncated response matrix for a variety of kernel magnitudes and truncation weights. We observed similar behavior for 14 and 17-qubit simulated experiments; readout error for quantum kernel experiments may be corrected reasonably well using a small fraction of the full bitflip response matrix.

Figure \ref{fig:bf_sim_ana} suggests a linear relationship between the kernel $K'$ computed in the presence of readout error and the noiseless kernel $K$. While readout error does not constitute a linear process in general (the underlying bitflip probabilities and bitstring distributions may be modified to give rise to arbitrary effects on $K$ within the bounds of Equation \ref{eq:bf_bounds}), by the arguments in Section \ref{app:circuit} demonstrating such an effect in general would imply minimal effect of readout error on the corresponding classifier. The degree to which readout error plays a role in quantum kernel methods is therefore an important research area for implementation on near-term processors.

\subsection{Crosstalk optimization}

Cross-talk between two-qubit gates on implemented on superconducting processors can contribute to decoherence and decrease circuit fidelity. Since our choice of circuit ansatz requires entire layers of entangling gates, we conducted diagnostic runs to determine whether executing these gates sequentially (in different staggering patterns) could improve performance compared to executing the gates simultaneously. We found that the completely parallel execution of entangling gates achieved the lowest cross entropy with respect to noiseless simulation compared to partially sequential arrangements. Therefore all entangling gates in this experiment were run in parallel.

 % % % % % % % % % % % % % % % % % % % % % % % % % % % % % % % % 
% \begin{center}
{\centering \section{Hardware error and performance}\label{sec:app_noise}}
% \end{center}

\begin{figure}[!htb]
	\centering
	\includegraphics[width=.9\textwidth]{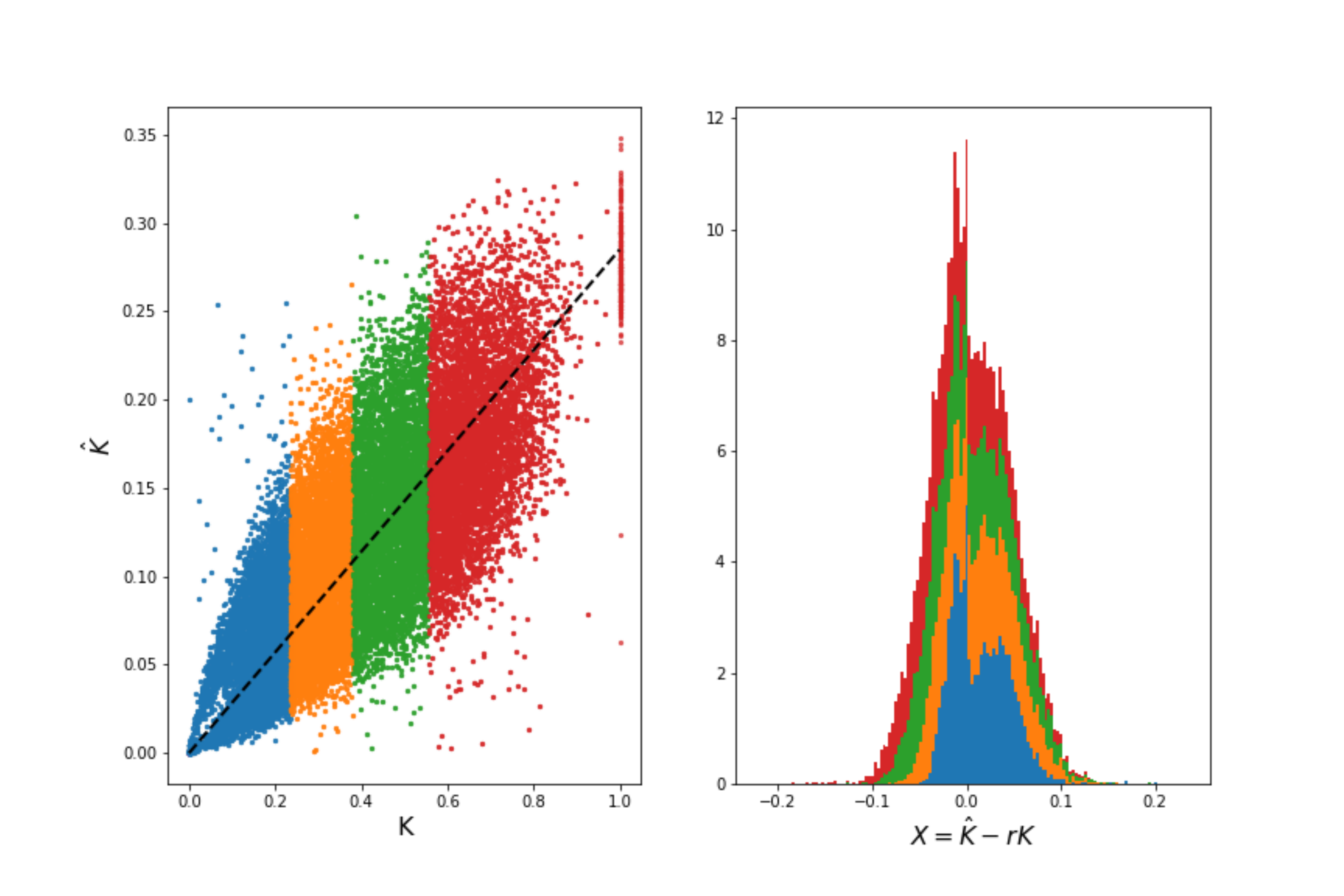}
	\caption{(left) Distribution of sampled $\hat{K}$ compared to simulated $K$ for 17 qubit train set ($m=210$) colored according to quartiles of $K$. The dashed line indicates the value of $rK$ for $r=0.29$, demonstrating that the hardware trends towards some scaled value of $K$. The mean value of $K_{ii}$ corresponding to $\Tr(|0\rangle \langle 0|)\equiv 1$ (elements in the top right) is a useful proxy for circuit fidelity and tends towards $30\%$ for the qubit counts investigated. (right) The distribution of $\hat{K}$ around $rK$ is irregular but exhibits consistent patterns with respect to kernel magnitude. The hardware error is therefore not normal with respect to (a scaled version of) $K$, which complicates bounds for guaranteed performance.}
	\label{fig:d11_corr}	
\end{figure}

Figure \ref{fig:d11_corr} shows a typical outcome for sampled kernel elements $\hat{K}_{ij}$ compared to their true values as determined from noiseless simulation. Notably, all of the hardware outcomes are strongly biased towards zero as a result of decoherence. The observation that the SVM decision function is scale-invariant (Appendix \ref{app:circuit}) suggests that the performance of an SVM trained using data subject to hardware error will only be affected to the degree that the sampled kernel elements $\hat{K}_{ij}$ differ from \textit{some linear transformation of the corresponding exact elements $K_{ij}$}, so that the circuit fidelity and other typical metrics for hardware performance are not predictive of classifier performance in any obvious way. For instance, we achieved comparable test accuracy to noiseless simulation for $n=14$ qubits despite circuit fidelity in the neighborhood of $30\%$.

\subsection{Effects of readout error in quantum kernel methods}

Since $\hat{K}$ is estimated by computing the empirical probability of the all zeros bitstring $p(0)$, the effects of readout error may be bounded in a straightforward manner. The following analysis will consider readout error as the sole source of noise and ignore statistical effects and other sources of decoherence. The resulting bounds apply to only to the infinite-shot limit but will be shown to be approximately correct in the low-shot limit. As in Section \ref{sec:readout_corr}, we let $p(y^n|x^n)$ describe the conditional probability for observing bitstring $y^n$ after exposing the bitstring $x^n$ to $n$ distinct bitflip channels, and let $q^k(y|x)$ for $x,y\in\{0,1\}$ describe the corresponding probability for observing "$y$" after exposing the $k$-th bit "$x$" to a single bitflip channel. After exposing the all-zeros bitstring to a stochastic bitflip process repeatedly, the lowest possible value for $\hat{K}$ occurs when no bitstrings transform into the all-zeros bitstring. The expected fraction of events remaining is  

\begin{equation}\label{eq:lb_p0}
    p(0|0) = \prod_k^n (1-q^k(1|0))
\end{equation}

The maximum increase in the observed $p(0)$ will result from transitions from other bitstrings into the all-zeros bistring. Intuitively this will occur almost entirely due to low-weight bitstrings with just a few bitflips. The probability of transition into the all-zeros bitstring from an arbitrary starting bitstring $y^n=y_1y_2\dots y_n$ is $p(0|y^n) = \prod_{k=1}^n q^k(0|y_k)$, while the log-odds of each term in this product may be rewritten as
\begin{align}
    \log q(0|y_k) = \log q(0|0) (1 - y_k) + \log q^k(0|1) y_k
\end{align}

The total log-odds for transition into $0$ is then
\begin{align}\label{eq:logodds}
    \log p(0|y^n) &= \sum_{k=1}^n \log q^k(0|0) + \sum_{k=1}^n \log \frac{q^k(0|1)}{q^k(0|0) } y_k
\end{align}
which is in the form $w \cdot y^n + b$ with $w \in \mathbb{R}^n, \, b\in \mathbb{R}$, indicating a linear dependence of $\log p(0|y^n)$ on $y^n$. Since the logarithm is strictly increasing, the corresponding integer programming problem for maximizing $p(0|y^n)$ is:

\begin{align}\label{eq:readout_maxprob}
    &\max_{w} \sum_{k=1}^n w_k x_k \\
    &\text{subject to } x_k \in \{0, 1\}, \qquad \sum_{k=1}^n x_k > 0
\end{align}

where the second constraint avoids the trivial solution of the all-zeros bitstring. By inspection of Equation \ref{eq:logodds}, $q^k(0|0) = 1 - q^k(1|0)$ implies that $w$ is strictly negative with the realistic assumption that $q^k(1|0) < 0.5, \, q^k(0|1) < 0.5$. In this case, the solution maximizing Equation \ref{eq:readout_maxprob} must be a bitstring with weight 1 with a ``1'' at the position $\text{argmax}_k q^k(0|1)$. Combined with Equation \ref{eq:lb_p0} this results in the bound:
\begin{equation}\label{eq:bf_bounds}
    \hat{K}\prod_k^n \left(1-q^k(1|0) \right)\leq \hat{K}' \leq \left(1-\hat{K}\right) \max_k q^k (0|1) + \hat{K}
\end{equation}
Where $\hat{K}'$ describes the estimated kernel element after readout error has occurred. This bound is plotted in Figure \ref{fig:bf_sim_ana}	for one instance of a 10-qubit readout error calibration. The form of these bounds further justify the need for large kernels, since the bound becomes increasingly loose as $\hat{K}$ approaches the magnitude of typical readout error probabilities on the quantum processor.

We now discuss the implementation of the readout error correction technique described in Appendix \ref{sec:readout_corr}. While routine calibration of the device returns diagnostic information on readout error probabilities $q^k(0|1)$ and $q^k(1|0)$, these quantities may drift in the time between calibration and experiment resulting in lower quality readout correction. To account for drift, we periodically estimated $q^k(0|1)$ and $q^k(1|0)$ by preparing a sequence of random bitstrings $|s\rangle$ and the complement $|s \oplus 1\rangle$ and then measured in the computational basis. We then computed empirical bitflip likelihoods for measurements in parallel over the specific qubits used in each experiment and computed the time-averaged likelihoods to use for readout correction. We averaged over the different prepared states to reduce the impact of imperfect state preparation on measured outcomes.

\begin{figure}[htp]
	\centering
	\includegraphics[width=.9\textwidth]{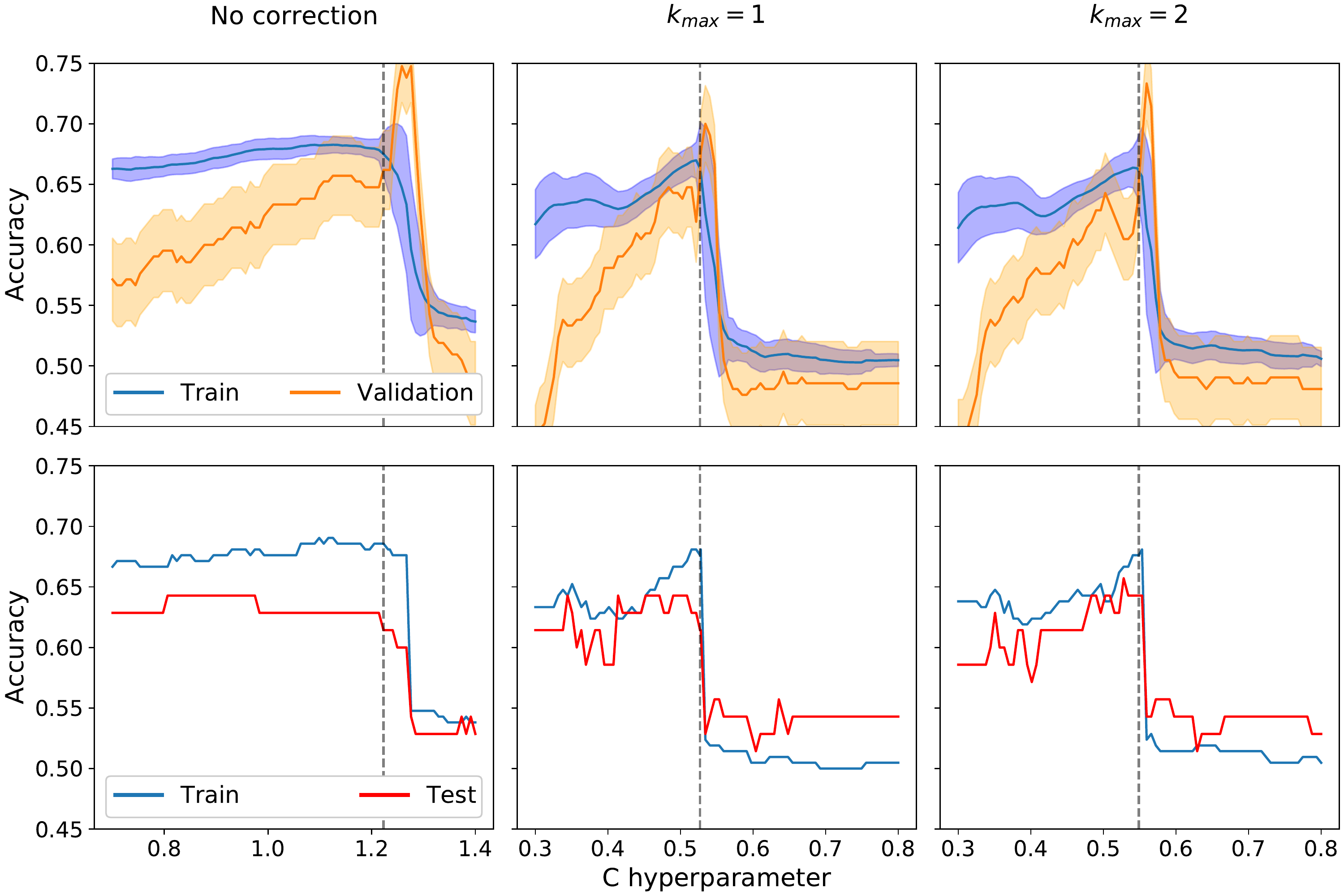}
	\caption{We analyzed the LOO cross-validation accuracy versus accuracy on the reserved test set to determine the effect of (truncated) readout correction using readout error likelihoods determined experimentally at periodic intervals during the 14-qubit experiment. While the trained classifiers are often able to achieve significantly higher accuracy on the reserved test set, the error-corrected validation accuracy is not predictive of improved test accuracy. For instance, the $k_{max}=2$ classifier achieves a 65.7\% validation accuracy and a 64.3\% test accuracy while the classifier with no readout correction achieves 66.1\% validation accuracy and 61.4\% test accuracy, indicating that the improved test score cannot be consistently predicted by LOOCV.}
	\label{fig:bf_hw_study}	
\end{figure}

Figure \ref{fig:bf_hw_study} shows learning curves computed for the 14-qubit experiment with and without readout correction for post-processing. While classifiers trained using error-corrected results are capable of achieving higher accuracies on the reserved test set, the choice of $C$ hyperparameter for improved test accuracy did not generally correspond to improved accuracies for the validation sets. Therefore, we could not consistently improve the classifier performance by applying readout correction and opted to present the results achieved without readout correction in Figure \ref{fig:hero1} in the main body. 

\subsection{Effects of statistical error on SVM accuracy}

While recent work \cite{liu2020rigorous} has established performance bounds relating SVM accuracy to statistical sampling error for quantum kernel methods, we conducted experiments using orders of magnitude fewer repetitions than necessary to achieve robust classification. In addition, modified SVM algorithms exist for processing noisy inputs in the data space \cite{bi2005support} but these modifications do not generalize to processing noise in the feature space (i.e. $\Delta k(x_i, x_j)$). Given these limitations, we chose to explore the effects of statistical noise numerically to determine the relative impact of statistical noise on final classifier accuracy compared to other sources of error.

\begin{figure}[htp]

\centering
\includegraphics[width=\textwidth]{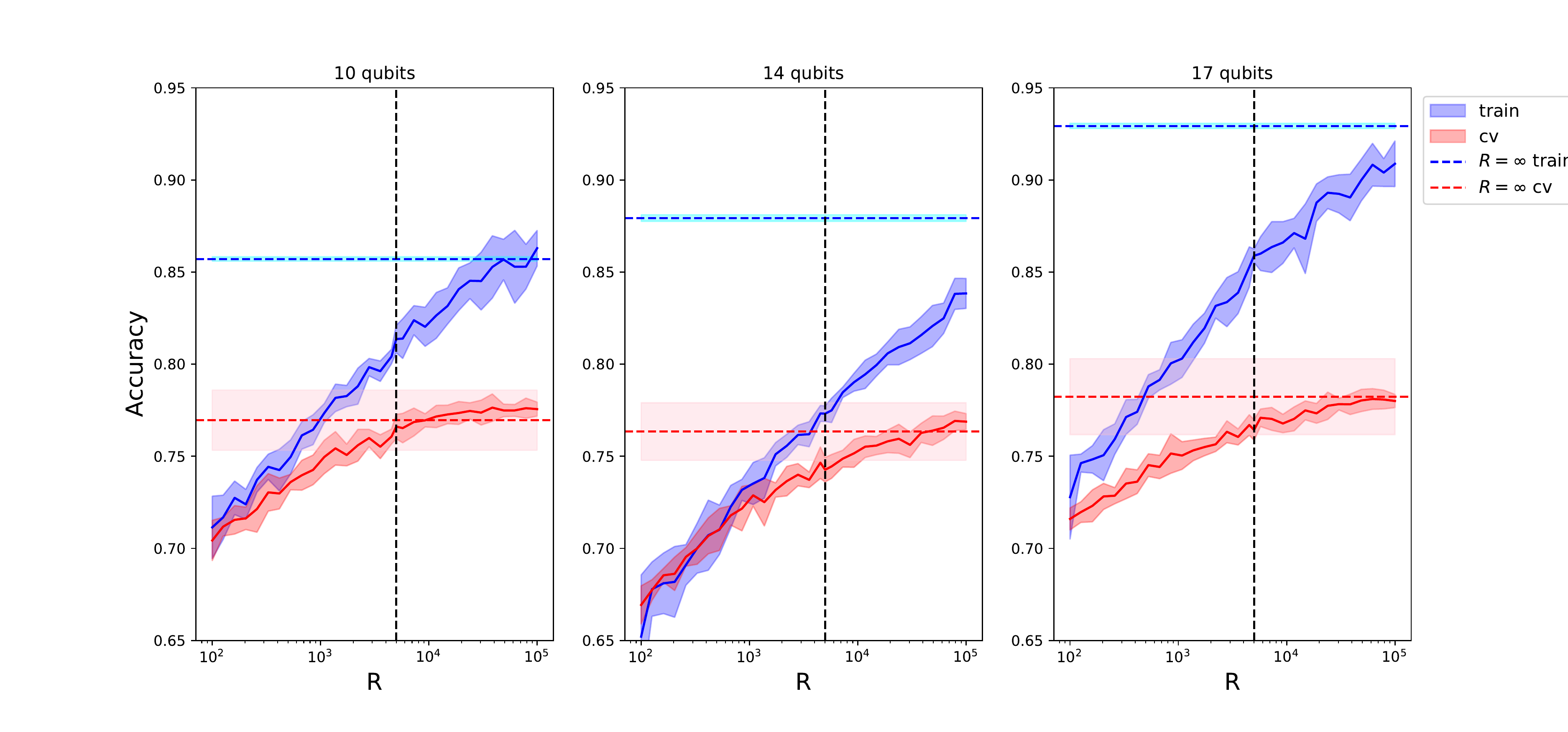}
\caption{We empirically investigated the effect of statistical noise by generating a sampling a binomial distribution with success probability equal to $K$ and enfocing symmetry on the resulting kernel matrix. The results show that additional circuit repetitions beyond $R\approx 5000$ provide diminishing returns to the validation accuracy of the classifier. We note that typically ~50,000 circuit repetitions per kernel element are required to achieve cross-validated accuracy comparable to noiseless simulation. Fill for finite-$R$ represents $1\sigma$ interval for  stratified 10-fold cross-validated train/test scores over 10 trials of downsampling to $R$ shots.}
\label{fig:stat_noise_emp}
\end{figure}

Figure \ref{fig:stat_noise_emp} shows simulated trials of the type 2 circuits used for experimental runs. Each circuit was initially simulated with an full wavefunction simulator, and then the amplitude $K = |\langle 0|\psi \rangle|^2$ was used to sample $R$ repetitions from the implied binomial distribution $\text{Bin}(R, K)$. The sampled kernel elements were used to train and validate SVM classifiers following the procedure outlined in Section \ref{sec:main_svm}. The results indicate there are diminishing returns in classifier accuracy beyond the $R=5000$ repetitions we used for the experiments, but that this choice incurs $\sim 1{\text -}2\%$ classifier error compared to the infinite-shot limit. As expected, greater statistical noise results in less overfitting for the model.

\subsection{Classifier results}

\begin{figure}[htp]

\centering
\includegraphics[width=.9\textwidth]{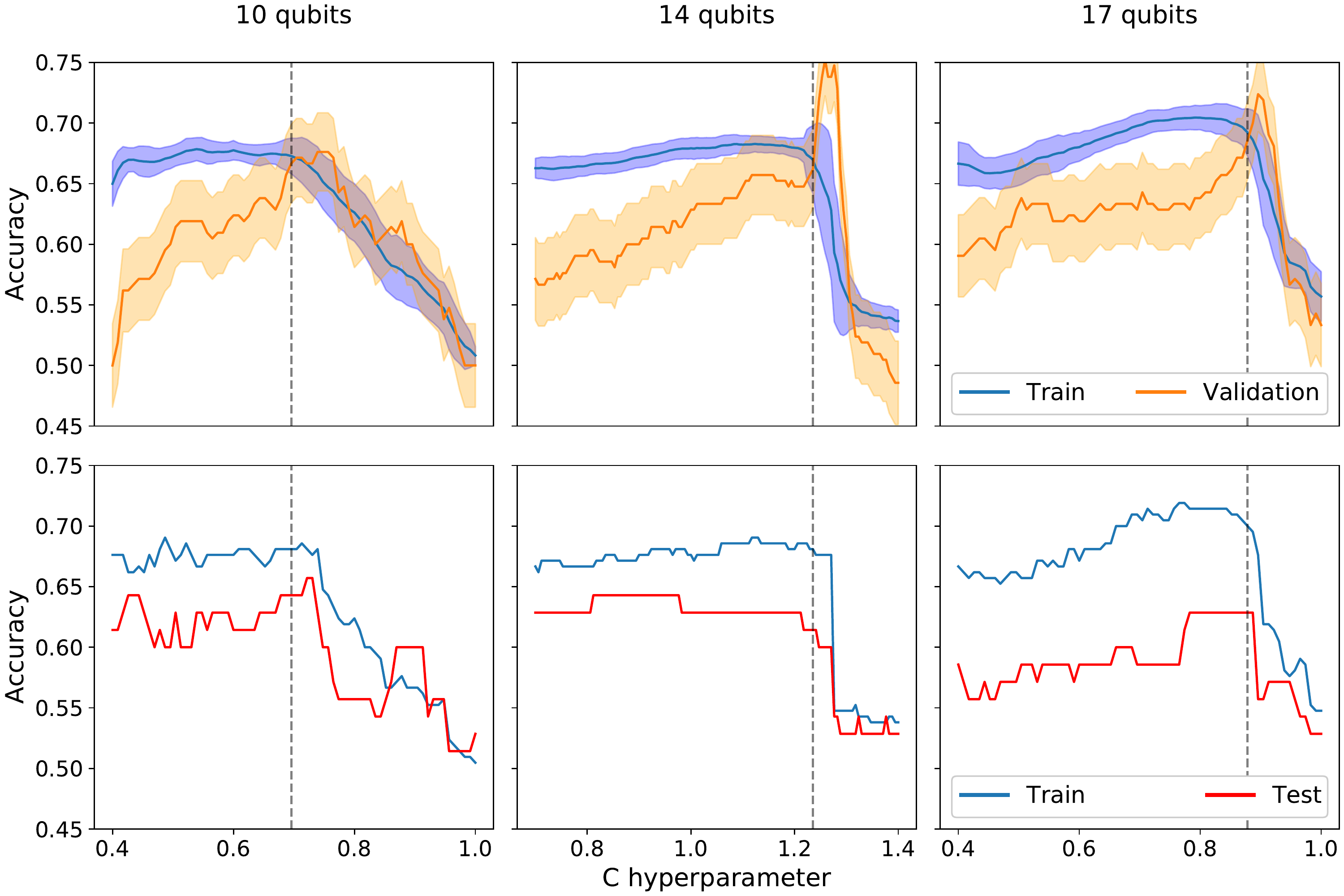}
\caption{Hyperparameter optimization for hardware kernels is performed by tuning the L1 penalty parameter $C$ via LOOCV on the training data (top row) during model validation, which then becomes fixed for evaluation of the model on the test set (bottom row). The capacity of the hardware-based models to overfit the data is drastically reduced, and oftentimes the SVM behavior becomes pathological. To avoid undesirable generalization behavior, the validation score corresponding to the optimal $C$ was required to be no greater than the corresponding training score. The vertical dashed line indicates the optimal $C$ decided in the validation stage.
% hep-qml/platanus_racemosa/hardware_scripts/20200803/big_run_analysis.ipynb
}
\label{fig:hw_c_reg_plots}
\end{figure}

Figure \ref{fig:hw_c_reg_plots} shows the result of tuning the hyperparameter $C$ controlling the $L2$ penalty for violating the SVM margin. The $C$ values resulting in optimal validation scores were then used to determine the final train/test scores reported in Figure \ref{fig:hero1} of the main body.

\end{document}